\documentclass[preprint,aps,nofootinbib]{revtex4}
\usepackage{dcolumn}
\usepackage{bm}
\usepackage{epsfig}
\usepackage{graphicx}
\usepackage{axodraw}
\usepackage{amsmath}
\usepackage{amssymb}
\usepackage{mathrsfs}
\usepackage{color}
\usepackage[usenames,dvipsnames]{xcolor}
\usepackage{hyperref}
\usepackage[caption=false]{subfig}

\def\fun#1#2{\lower3.6pt\vbox{\baselineskip0pt\lineskip.9pt
  \ialign{$\mathsurround=0pt#1\hfil##\hfil$\crcr#2\crcr\sim\crcr}}}

\def\simlt{\stackrel{<}{{}_\sim}}
\def\simgt{\stackrel{>}{{}_\sim}}

\input epsf

\newenvironment{Eqnarray}%
         {\arraycolsep 0.14em\begin{eqnarray}}{\end{eqnarray}}
\newcommand{\be}{\begin{equation}}
\newcommand{\ee}{\end{equation}}
\newcommand{\bea}{\begin{Eqnarray}}
\newcommand{\eea}{\end{Eqnarray}}

\def\nn{\nonumber}
\def\half{\tfrac{1}{2}}
\def\phm{\phantom{-}}
\def\sinbii{s^2_\beta}
\def\sinbiv{s^4_\beta}
\def\cosb{c_\beta}
\def\sinb{s_\beta}
\def\cosa{c_\alpha}
\def\sina{s_\alpha}
\def\cosbii{c^2_\beta}
\def\cosbiv{c^4_\beta}
\def\cbma{c_{\beta-\alpha}}
\def\sbma{s_{\beta-\alpha}}
\def\lsim{\mathrel{\raise.3ex\hbox{$<$\kern-.75em\lower1ex\hbox{$\sim$}}}}
\def\gsim{\mathrel{\raise.3ex\hbox{$>$\kern-.75em\lower1ex\hbox{$\sim$}}}}
\def\lsub#1{_{\lower 1.5pt\hbox{$\scriptstyle#1$}}}
\def\Ref#1{Ref.~\cite{#1}}

\def\eq#1{Eq.~(\ref{#1})}
\def\eqs#1#2{Eqs.~(\ref{#1}) and (\ref{#2})}

\def\eqst#1#2{Eqs.~(\ref{#1})--(\ref{#2})}
\def\eqthree#1#2#3{Eqs.~(\ref{#1}), (\ref{#2}) and (\ref{#3})}
\def\Eq#1{Eq.~(\ref{#1})}

\def\Eqs#1#2{Eqs.~(\ref{#1}) and (\ref{#2})}

\begin{document}

\begin{flushright}
\vspace*{-0.5cm}
\vspace{-0.4cm}EFI-14-36, FERMILAB-PUB-14-392-T, MCTP-14-37, SCIPP 14/16\\
\end{flushright}

\vspace*{1.8cm}

\title{Complementarity Between Non-Standard Higgs Searches  and Precision Higgs Measurements in the MSSM}

\vspace*{-0.2cm}

\author{
\vspace{0.1cm}
\mbox{\bf Marcela Carena$^{\,a,b,c}$, Howard E. Haber$^{\,d,e}$, Ian Low$^{\,f,g}$,} \\
\mbox{\bf Nausheen R. Shah$\,^{h}$, and Carlos E.~M.~Wagner$^{\,b,c,f}$}
 }
\affiliation{
\vspace*{.1cm}
$^a$\mbox{\footnotesize{Fermi National Accelerator Laboratory, P.O. Box 500, Batavia, IL 60510}}\\
$^b$\mbox{\footnotesize{Enrico Fermi Institute, University of Chicago, Chicago, IL 60637}}\\
$^c$\mbox{\footnotesize{Kavli Institute for Cosmological Physics, University of Chicago, Chicago, IL 60637}}\\
$^d$\mbox{\footnotesize{Santa Cruz Institute for Particle Physics, University of California, Santa Cruz,  CA 95064}} \\
$^e$\mbox{\footnotesize{Ernest Orlando Lawrence Berkeley National Laboratory,
University of California, Berkeley, CA 9472}}\\
$^f$\mbox{\footnotesize{High Energy Physics Division, Argonne National Laboratory, Argonne, IL 60439}}\\
$^g$\mbox{\footnotesize{Department of Physics and Astronomy, Northwestern University, Evanston, IL 60208}} \\
$^h$\mbox{\footnotesize{Michigan Center for Theoretical Physics,
University of Michigan, Ann Arbor, MI 48109}}\\
}

\vskip -0.5in

\begin{abstract}
%

Precision measurements of the Higgs boson properties at the LHC provide relevant constraints on
possible weak-scale extensions of the Standard Model (SM). In the context of the Minimal Supersymmetric Standard Model (MSSM)
these constraints seem to suggest that all the additional, non-SM-like Higgs bosons  should be
heavy, with masses larger than about 400 GeV. This article shows that such results do not hold
 when the theory approaches the conditions for ``alignment independent of decoupling'', 
 where  the lightest CP-even Higgs boson has SM-like tree-level couplings to fermions and gauge bosons, independently of the non-standard Higgs boson masses.
 The combination of current bounds from direct Higgs boson searches at the LHC, along with the 
alignment conditions, have a significant impact on the allowed MSSM parameter space yielding
light additional Higgs bosons. In particular, after ensuring the correct mass for the lightest CP-even
Higgs boson, we find that precision measurements and direct searches are complementary, and may
soon be able to probe the region of non-SM-like Higgs boson with masses below the top quark pair mass
threshold of 350 GeV and low to moderate values of $\tan\beta$.

\end{abstract}
\thispagestyle{empty}

\maketitle

 \section{Introduction}

The recent discovery of a scalar resonance at the LHC, with a mass of about 125~GeV and properties resembling that of the Higgs boson of the Standard Model (SM)~\cite{Aad:2012tfa,Chatrchyan:2012ufa}, has revived interest in particle physics models in which a SM-like Higgs boson arises in a natural way. The Minimal Supersymmetric extension of the SM (MSSM)
is an example of such a model~\cite{mssmhiggs,Gunion:1984yn,mssmhiggsreview1,mssmhiggsreview2}.  The Higgs sector of the MSSM consists of two Higgs doublets with tree-level quartic couplings which are related to the squares of the weak gauge couplings. The tree-level Higgs boson mass spectrum consists of two neutral CP-even Higgs scalars, $h$ and $H$ (with $m_h\leq m_H$), a CP-odd scalar, $A$, and a charged Higgs pair, $H^\pm$. The quartic scalar couplings receive quantum corrections whose leading contributions are proportional to the fourth power of the top-quark Yukawa coupling~\cite{mssmhiggsradcorr}.  For top squark masses below a few TeV, an upper bound on  the lightest CP-even Higgs boson mass of about 135~GeV is obtained~\cite{mssmhiggsupperbound}.\footnote{The same upper bound is obtained in the presence of explicit CP-violating phases in the supersymmetry breaking mass parameters, which affect the Higgs sector via radiative corrections. In this paper, we will simplify our analysis by neglecting these CP-violating phases, in which case the neutral Higgs bosons of the MSSM are CP eigenstates~\cite{cpviolatingmssmhiggs}.}
The observed Higgs boson mass is comfortably below this predicted upper bound.

For large values of the supersymmetric particle masses, the properties of $h$ are determined by $m_A$ and the third generation supersymmetric spectrum that governs the size of the quantum corrections to the quartic couplings. When $m_A\gg m_h$, one finds that $m_H\sim m_A\sim m_{H^\pm}$, with corresponding squared-mass differences of $\mathcal{O}(m_Z^2)$. Hence, all non-standard Higgs bosons are heavy and decouple from the low-energy effective theory at the weak scale, which then naturally consists of the light CP-even Higgs boson, $h$, with SM-like couplings, as suggested by current measurements. This is the well known decoupling limit of the MSSM Higgs sector.

In contrast, for values of $m_A\sim\mathcal{O}(m_h)$, the coupling of $h$ to bottom-quark pairs tends to be enhanced with respect to the SM value. Since the coupling to bottom-quarks controls the width of the Higgs boson, such an enhancement leads to an increase of the Higgs width and therefore a reduction of the branching ratios of the Higgs decay into neutral and charged gauge bosons.  Such a reduction can become significant for values of $m_A$  below 300~GeV.  Hence, precision studies of the lightest CP-even Higgs boson properties can lead to significant constraints on the allowed parameter space of the theory.  The large increase of the Higgs boson width may be avoided if the properties of $h$ are SM-like, which can occur either via the decoupling limit~\cite{Gunion:2002zf,whitepaper,Haber:2013mia} or the so-called {\em alignment} limit \cite{Craig:2013hca,whitepaper,Carena:2013ooa,Haber:2013mia}.

The alignment limit arises when one of the CP-even Higgs bosons, when expressed as a linear combination of the real parts of the two neutral Higgs fields, lies in the same direction in the two Higgs doublet field space as the two neutral Higgs vacuum expectation values.  This alignment does not in general depend on the masses of the non-standard Higgs bosons. In the MSSM the alignment limit arises due to an accidental cancellation, i.e. not due to any of the usual symmetries of the MSSM,  between tree-level and loop-corrected effects resulting from new structures in the potential that are absent at tree-level~\cite{Carena:2013ooa}.  However, this cancellation occurs quite generically for some value of the ratio of neutral Higgs vacuum expectation values, $\tan\beta$, which depends critically on $\mu$, the supersymmetric Higgs mass in the potential, and $A_t$, the stop mixing parameter. In particular, alignment at lower values of $\tan\beta$ typically requires $\mu$ and $A_t$ to be larger than the characteristic mass scale for the top squarks~\cite{Carena:2013ooa,CHLM}, leading to important phenomenological constraints in the MSSM.

One can also search directly for the heavier Higgs bosons of the MSSM at the LHC. The most sensitive search channel is associated with the neutral Higgs boson decays into $\tau^+\tau^-$, produced in gluon fusion processes or in association with $b$-quarks.  This channels becomes particularly sensitive for low values of the heavier Higgs boson masses and large values of $\tan\beta$, and allows one to set a bound on $m_A$ that extends from 200~GeV at values of $\tan\beta\sim 10$, up to 900~GeV for $\tan\beta\sim 50$.  Lower values of $\tan\beta$ in the range $3\lsim\tan\beta\lsim 10$, still consistent with the observed mass of the lightest CP-even Higgs mass for stop masses below a few TeV, remain mostly unconstrained  by these searches, due to a suppression of the production cross-section times the Higgs decay branching ratio into $\tau^+\tau^-$.   This branching ratio depends on  possible decays into both non-supersymetric and supersymmetric final states~(e.g. neutralino and chargino pairs). The latter are suppressed for large values of $\mu$, for which alignment is obtained. Therefore there is an interesting correlation between the properties of the lightest CP-even Higgs boson and the rate of non-standard Higgs boson decays into the $\tau^+\tau^-$ channel.

In this paper we shall discuss the complementarity of precision studies of the lightest CP-even Higgs boson and the search for heavier neutral Higgs bosons in the $\tau^+\tau^-$ channel. In particular, since we assume the lightest CP-even Higgs is the one discovered at around 125 GeV, we will design our benchmarks in such a way that the correct mass is obtained for $h$ over the entire $m_A$--$\tan\beta$ plane, in contrast to previously established benchmarks. This is an especially important point when considering properties of  $h$ where its mass plays an essential role. The lightest CP-even Higgs mass is also relevant in the determination of the decay branching fractions of $H$ and $A$, since the decay modes $H \to hh$ and $A \to hZ$ become important at low values of $\tan\beta$ and their rates depend crucially on $m_h$.

This paper is organized as follows. In Section~\ref{sec:MSSMHiggs} we present an overview of the two Higgs doublet model (2HDM)\footnote{For a review of the two Higgs doublet model see, e.g., Refs.~\cite{Gunion:1989we} and \cite{Branco:2011iw}.} and its application to the Higgs sector of the MSSM, with emphasis on the behavior of the down-type quark couplings to the lightest CP-even Higgs boson and the associated condition of alignment at large values of $\mu$ and $A_t$.  In Section III we discuss the constraints on $m_A$ that come from the precision study of the lightest CP-even Higgs boson properties for different values of $\mu$.  In Section IV we analyze the sensitivity of the non-standard Higgs searches on the value of the $\mu$ parameter, and compare it with the results obtained in Section III. We reserve Section V for our conclusions.  A detailed description of our interpretation of the experimental limits presented by CMS for the direct searches of $H$ and $A$ is presented in Appendix A.  Finally, the comparison of the $hVV$ ($VV=W^+ W^-$ or $ZZ$) and $h\gamma\gamma$ couplings is provided in Appendix B.

 \section{Overview of the MSSM Higgs Sector} \label{sec:MSSMHiggs}

\subsection{The Two Higgs Doublet Model  (2HDM): Theoretical Background}

The scalar potential of the most general two-Higgs-doublet extension of the SM may be written in terms of two Higgs doublet fields, $\Phi_i$ ($i=1,2$), each carrying the same hypercharge quantum number, $Y_H = \half$~\cite{Haber:1993an}:
\bea
\label{eq:generalpotential}
V &=& m_{11}^2 \Phi_1^\dagger \Phi_1+m_{22}^2 \Phi_2^\dagger \Phi_2-m_{12}^2 (\Phi_1^\dagger \Phi_2 +{\rm h.c.}) +\tfrac12 \lambda_1 ( \Phi_1^\dagger \Phi_1)^2+\tfrac12 \lambda_2 ( \Phi_2^\dagger \Phi_2)^2 \nonumber \\
&& +\lambda_3 ( \Phi_1^\dagger \Phi_1)( \Phi_2^\dagger \Phi_2)+\lambda_4 ( \Phi_1^\dagger \Phi_2)( \Phi_2^\dagger \Phi_1) \nonumber \\
&&+ \left\{ \tfrac12 \lambda_5 ( \Phi_1^\dagger \Phi_2)^2 + [ \lambda_6 (\Phi^\dagger_1\Phi_1)+ \lambda_7 (\Phi^\dagger_2\Phi_2)]\Phi_1^\dagger\Phi_2 + {\rm h.c.} \right\}\ ,
\eea
where $m_{11}^2$, $m_{22}^2$ and $\lambda_1,\ldots,\lambda_4$ are real parameters and $m_{12}^2$, $\lambda_5$, $\lambda_6$ and $\lambda_7$ are potentially complex.  For simplicity, we shall assume that the scalar potential is explicitly CP conserving, in which case we can assume, without loss of generality, that all scalar potential parameters are real.

We parameterize the scalar doublets in terms of a complex charged field and two neutral real fields,
\be \label{Phisubi}
\Phi_i = \left( \begin{array}{c}
       \phi_i^+ \\
       \frac1{\sqrt{2}}(v_i+\phi_i^0+i a_i^0)
       \end{array} \right) \,,
 \ee
where the minimum of the scalar potential is at
\be
\langle \Phi_i \rangle = \frac1{\sqrt{2}}  \left( \begin{array}{c}
        0\\
        v_i
         \end{array} \right) \ ,
\ee
and
\be
v\equiv \sqrt{|v_1|^2+|v_2|^2} \simeq 246 \ {\rm GeV}\,.
\ee
Since the scalar potential and the vacuum preserve CP, there exists a basis of scalar fields where all scalar potential parameters, as well as $v_1$ and $v_2$, are real and non-negative. Therefore, one can define
\be
t_\beta \equiv \tan\beta = \frac{v_2}{v_1} \,,
\ee
where $0\leq\beta\leq\half\pi$.

The squared-mass matrix for the CP-even scalars can be expressed as~\cite{Gunion:2002zf}
\be
\label{eq:cpemass}
\mathcal{M}^2  = \left(\begin{array}{cc}
                                                    \mathcal{M}_{11}^2 &\quad  \mathcal{M}_{12}^2 \\
                                                     \mathcal{M}_{12}^2 &\quad  \mathcal{M}_{22}^2
                                                     \end{array}\right)
                                                     \equiv
                                                     m_A^2 \left(\begin{array}{cc}
                                                    s_\beta^2 &\,\,\, -s_\beta c_\beta \\
                                                     -s_\beta c_\beta &\,\,\, c_\beta^2
                                                     \end{array}\right)
                                 + v^2    \left(\begin{array}{cc}
                                                            L_{11} & \,\,\,L_{12} \\
                                                            L_{12} & \,\,\,L_{22}
                                                            \end{array}\right) \ ,
\ee
where  $s_\beta\equiv\sin\beta=v_2/v$, $c_\beta\equiv\cos\beta=v_1/v$,
\be \label{Amass}
m_A^2=m_{12}^2-\half v^2(2\lambda_5+\lambda_6 t_\beta^{-1}+\lambda_7 t_\beta)\,,
\ee
is the squared-mass of the CP-odd Higgs boson and
\bea
    L_{11} &=& \lambda_1 c_\beta^2 +2\lambda_6 s_\beta c_\beta+\lambda_5 s_\beta^2 \ , \label{eqL11}\\
    L_{12}&=& (\lambda_3+\lambda_4)s_\beta c_\beta +\lambda_6 c_\beta^2 +\lambda_7 s_\beta^2\ ,\label{eqL12}
 \\
     L_{22}&=&  \lambda_2 s_\beta^2 +2\lambda_7 s_\beta c_\beta+\lambda_5 c_\beta^2 \ .
     \label{eqL22}
 \eea

Diagonalizing the squared-mass matrix, $\mathcal{M}^2$, given in \eq{eq:cpemass} yields
two CP-even Higgs mass eigenstates, $h$ and $H$, with squared-masses
\be \label{cpmasses}
m^2_{H,h}=\half\bigl[\mathcal{M}^2_{11}+\mathcal{M}^2_{22}\pm\Delta\bigr]\,,
\ee
where $m_h\leq m_H$ and the non-negative quantity $\Delta$ is defined by
\be
\Delta\equiv\sqrt{(\mathcal{M}^2_{11}-\mathcal{M}^2_{22})^2+4(\mathcal{M}^2_{12})^2}\,.
\ee
In particular, $m_h^2\leq\mathcal{M}^2_{ii}\leq m^2_H$, $i=1,2$.
We also note that the two equations,
\be
{\rm Tr}~\mathcal{M}^2=m_H^2+m_h^2\,,\qquad\quad
{\rm det}~\mathcal{M}^2=m_H^2 m_h^2\,,
\ee
yield the following result:
\be
|\mathcal{M}^2_{12}|
=\sqrt{(m_H^2-\mathcal{M}^2_{11})(\mathcal{M}^2_{11}-m_h^2)}
=\sqrt{(\mathcal{M}^2_{22}-m_h^2)(\mathcal{M}^2_{11}-m_h^2)}
\,.\label{relation}
\ee

The CP-even Higgs mass-eigenstate fields can be expressed in terms of the neutral scalar fields, $\phi_1^0$ and $\phi_2^0$, defined in \eq{Phisubi},
\be\label{ralpha}
\left(\begin{array}{c}
                    H\\
                    h
        \end{array}\right) =
              \left(\begin{array}{cc}
                    \phm c_\alpha & \,\,\,s_\alpha \\
                    -s_\alpha & \,\,\,c_\alpha
        \end{array}\right)
        \left(\begin{array}{c}
                    \phi_1^0\\
                    \phi_2^0
        \end{array}\right)  \,,
\ee
where the mixing angle $\alpha$ is defined modulo $\pi$, $c_\alpha\equiv \cos\alpha$ and $s_\alpha\equiv \sin\alpha$.  It is often convenient to restrict the range of the mixing angle to $|\alpha|\leq\half\pi$.  In this case, $c_\alpha$ is non-negative and is given by
\be \label{ca}
c_\alpha=\sqrt{\frac{\Delta+\mathcal{M}^2_{11}-\mathcal{M}^2_{22}}{2\Delta}}\,,
\ee
and the sign of $s_\alpha$ is given by the sign of $\mathcal{M}_{12}^2$.  Explicitly, we have
\be \label{sa}
s_\alpha=\frac{\sqrt{2}\,\mathcal{M}_{12}^2}{\sqrt{\Delta(\Delta+\mathcal{M}^2_{11}-\mathcal{M}^2_{22})}}.
\ee
Using \eqs{cpmasses}{relation}, one can derive alternative forms for \eqs{ca}{sa},
\be \label{csalt}
c_\alpha=\sqrt{\frac{\mathcal{M}^2_{11}-m_h^2}{m_H^2-m_h^2}}\,,\qquad\quad
s_\alpha={\rm sgn}(\mathcal{M}^2_{12})\sqrt{\frac{m_H^2-\mathcal{M}^2_{11}}{m_H^2-m_h^2}}\,.
\ee
For completeness, we also record the squared mass of the charged Higgs boson, $H^\pm$,
\be
m_{H^\pm}^2= m_A^2+\half v^2(\lambda_5-\lambda_4)\,,
\ee
where $m^2_A$ is given by \eq{Amass}.

The recently discovered Higgs boson, exhibits couplings to gauge bosons and fermions that are consistent (within experimental errors) with SM expectations.  If the 2HDM is realized in nature, it is tempting to identify the observed Higgs boson with the lightest CP-even scalar, $h$, which is a linear combination of $\phi^0_1$ and $\phi^0_2$ as indicated in \eq{ralpha}. If $h$ is SM-like, then it follows that in the $\phi^0_1$--$\phi^0_2$ field space, $h$ points roughly in a direction parallel to the direction of the scalar field vacuum expectation values. The implications of this observation will now be examined in more detail.

Since the Higgs couplings to gauge bosons are more accurately measured, we first focus on these. The tree-level coupling of $h$ to $VV$ (where $VV=W^+W^-$ or $ZZ$), normalized to the corresponding SM coupling, is given by
\be \label{hvvcoup}
g_{hVV}=g\lsub{hVV}^{\rm SM}~s_{\beta-\alpha}\,.
\ee
Thus, if the $hVV$ coupling is SM-like, it follows that
\be \label{cbmasmall}
|c_{\beta-\alpha}|\ll 1\, ,
\ee
where $\cbma\equiv\cos(\beta-\alpha)$ and $\sbma\equiv\sin(\beta-\alpha)$. It is therefore instructive to consider under what conditions \eq{cbmasmall} can be achieved.

At this stage, there is nothing that distinguishes the Higgs doublets, since one is free to construct new doublet fields that are linear combinations of $\Phi_1$ and $\Phi_2$~\cite{Davidson:2005cw}.  Consequently, the parameters $\alpha$ and $\beta$ are not physical, although the quantity $(\beta-\alpha)$ is physical (modulo~$\pi$) since it is related to an observable coupling.  To derive an explicit formula for $c_{\beta-\alpha}$, it is convenient to define the so-called \textit{Higgs basis} of scalar doublet fields~\cite{higgsbasis,Branco:1999fs},
\be \label{cphiggsbasisfields}
H_1=\begin{pmatrix}H_1^+\\ H_1^0\end{pmatrix}\equiv \frac{v_1 \Phi_1+v_2\Phi_2}{v}\,,
\qquad\quad H_2=\begin{pmatrix} H_2^+\\ H_2^0\end{pmatrix}\equiv\frac{-v_2 \Phi_1+v_1\Phi_2}{v}
 \,,
\ee
so that $\langle{H_1^0}\rangle=v/\sqrt{2}$ and $\langle{H_2^0}\rangle=0$. From this one can immediately identify that the scalar doublet $H_1$ is the one that will have SM tree-level couplings to all the SM particles. It follows that if one of the CP-even neutral Higgs mass eigenstates is SM-like, then it must be approximately aligned with the real part of the neutral field $H_1^0$.

The scalar potential, when expressed in terms of the doublet fields, $H_1$ and $H_2$, has the same form as \eq{eq:generalpotential}, but now with coefficients $\lambda_i \rightarrow Z_i$. Indeed, one can translate all the formulae obtained previously in the original basis of the scalar fields, $\{\Phi_1\,,\,\Phi_2\}$, into the Higgs basis by taking $\beta\to 0$ and $\alpha\to (\alpha-\beta)$. Hence, in the limit of $ c_{\beta-\alpha}\to 0$ we have $h\simeq[ \sqrt{2}\,{\rm Re} ~\!(H_1^0)-v]$, which means that $h$ is aligned with the real part of the neutral component of the Higgs basis field that possesses the non-zero vacuum expectation value. The existence of a neutral scalar mass-eigenstate with the properties of the SM Higgs boson is equivalent to demanding that $c_{\beta-\alpha}=0$.

The scalar potential in the Higgs basis is given by,
\be \label{higgsbasispot}
\mathcal{V} \supset  \ldots+\half Z_1(H_1^\dagger H_1)^2+\ldots +\big[Z_5(H_1^\dagger H_2)^2+Z_6 (H_1^\dagger
H_1) H_1^\dagger H_2+{\rm
h.c.}\bigr]+\ldots\,,
\ee
where~\cite{Gunion:2002zf,Davidson:2005cw}
\bea
Z_1 & \equiv & \lambda_1\cosbiv+\lambda_2\sinbiv+\half(\lambda_3+\lambda_4+\lambda_5)s^2_{2\beta}
+2s_{2\beta}\bigl[\cosbii\lambda_6+\sinbii\lambda_7\bigr]\,,\label{zeeone}\\
Z_5 & \equiv & \tfrac{1}{4} s^2_{2\beta}\bigl[\lambda_1+\lambda_2-2(\lambda_3+\lambda_4+\lambda_5)\bigr]+\lambda_5-s_{2\beta}c_{2\beta}(\lambda_6-\lambda_7)\,,\label{zeefive}\\
Z_6 & \equiv & -\half s_{2\beta}\bigl[\lambda_1\cosbii-\lambda_2\sinbii-(\lambda_3+\lambda_4+\lambda_5)c_{2\beta}\bigr]+\cosb c_{3\beta}\lambda_6+\sinb s_{3\beta}\lambda_7\,,\label{zeesix}
\eea
and the shorthand notation, $s_{2\beta}\equiv\sin 2\beta$, $c_{2\beta}\equiv\cos 2\beta$, etc.,~has been employed.

It is straightforward to compute the  CP-even Higgs squared-mass matrix in the Higgs basis,
\be
\mathcal{M}_H^2=\begin{pmatrix} Z_1 v^2 & \quad Z_6 v^2 \\  Z_6 v^2 & \quad m_A^2+Z_5 v^2\end{pmatrix}\,.
\ee
The significance of $Z_1$ and $Z_6$ can now be immediately discerned. The upper diagonal element of the squared-mass matrix in the Higgs basis, $\mathcal{M}_{H11}^2=Z_1 v^2$, implies that $m_h^2\leq Z_1 v^2$, whereas the off-diagonal element, $\mathcal{M}_{H12}^2= Z_6 v^2$, governs the mixing between the Higgs basis fields $H_1^0$ and $H_2^0$. The presence of this mixing yields a non-alignment of the mass eigenstates, $h$ and~$H$, from the neutral Higgs basis states, $H_1^0$ and $H_2^0$.  Moreover, if $|Z_6|\ll 1$, then the mass eigenstate approximately aligned with ${\rm Re}~\!(H_1^0)$ behaves like the SM Higgs boson. Alternatively, if $m_A^2\gg Z_i v^2$ ($i=1,5,6$), then $Z_1$ and $Z_6$ can be treated as small perturbations in the diagonalization of the CP-even Higgs squared-mass matrix, $h$ is again SM-like, since it is approximately aligned with ${\rm Re}~\!(H_1^0)$.

The mixing angle in the Higgs basis can be obtained simply by using the relations written down for the original basis of the scalar fields. Translating our previous results into the Higgs basis
by taking $\alpha\to\alpha-\beta$, $\mathcal{M}_{11}^2\to Z_1 v^2$ and $\mathcal{M}_{12}^2\to Z_6 v^2$, \eq{relation} implies that
\bea \label{zsix}
|Z_6|v^2=\sqrt{(Z_1 v^2-m_h^2)(m_H^2-Z_1 v^2)}\,,
\eea
and \eq{csalt} yields,
\be \label{sc}
c_{\beta-\alpha}=\sqrt{\frac{Z_1 v^2-m_h^2}{m_H^2-m_h^2}}\,,\qquad\quad
s_{\beta-\alpha}=-{\rm sgn}\!\left(Z_6\right)\sqrt{\frac{m_H^2-Z_1 v^2}{m_H^2-m_h^2}}\,,
\ee
in a convention where $|\beta-\alpha|\leq\half\pi$. Actually, it is somewhat more convenient to adopt a different sign convention in which $s_{\beta-\alpha}$ is non-negative and the sign of $c_{\beta-\alpha}$ is fixed by $Z_6$, since in this convention the sign of the $hVV$ coupling is the same as in the SM [cf.~\eq{hvvcoup}].  In particular, if we assume that $0\leq\beta-\alpha\leq\pi$, then we can use \eqs{zsix}{sc} rewrite $\cbma$ in the more useful form,
\be
c_{\beta-\alpha}=\frac{-Z_6 v^2}{\sqrt{(m_H^2-m_h^2)(m_H^2-Z_1 v^2)}}\,.\label{cbmaf}
\ee

Tree-level unitarity (or perturbativity) constraints yield upper limits on the quartic scalar coupling parameters that are roughly of the form $\lambda_i/(4\pi)\lsim 1$, with similar limits applying to $Z_1$ and $Z_6$. In light of these constraints, there are two ways to achieve $|c_{\beta-\alpha}|\ll 1$, corresponding to alignment and hence to a SM-like $h$.

First, if $m_H^2 \gg m_h^2$, $Z_1v^2$, $Z_6 v^2$, then it follows that
\be \label{dl}
c_{\beta-\alpha}\sim\mathcal{O}\left(\frac{Z_6 v^2}{m_H^2}\right)\,,
\qquad\quad
Z_1 v^2- m_h^2\sim\mathcal{O}\left(\frac{ Z_6^2 v^4}{m_H^2} \right)\,.
\ee
This is the well-known decoupling limit~\cite{Gunion:2002zf}, in which alignment is achieved when $m_H$, $m_A$, $m_{H^\pm}\gg m_h$. Integrating out the heavy scalars yields an effective theory with one CP-even scalar, $h$, with SM couplings.

In contrast, suppose that $|Z_6|\ll 1$. This is the only case that can result in exact alignment (corresponding to $Z_6=0$), and we will henceforth refer to this case as the \textit{alignment limit}, which exists independently of the decoupling limit.  Indeed, \eqs{zsix}{cbmaf} imply that if $|Z_6|\ll 1$ and $m_h^2\simeq Z_1 v^2$ then,
\be
\label{alignmentconditions}
c_{\beta-\alpha}\sim\mathcal{O}(Z_6)\,,\qquad\quad Z_1 v^2-m_h^2\sim\mathcal{O}(Z_6^2 v^2)\,,
\ee
in which case $h$ is SM-like.\footnote{If $|Z_6|\ll 1$ and $m_H\simeq Z_1 v^2$, then $s_{\beta-\alpha}\ll 1$, and we would identify the SM-like Higgs boson with~$H$. This possibility cannot be completely ruled out for a general 2HDM but is very unlikely in the MSSM Higgs sector.} Note that the alignment limit can be achieved even in a case where $m_H \sim \mathcal{O}(v)$.

To make contact with the results of Ref.~\cite{Carena:2013ooa}, one can compute $c_{\beta-\alpha}=(\cosb\cosa+\sinb\sina)$ using \eqs{relation}{csalt}. Additional simplification can be implemented by noting that $\mathcal{M}_{11}^2+\mathcal{M}_{22}^2=\Delta+2m_h^2$, which allows us to remove $\Delta$ in favor of $m_h^2$.  The end result is
\be \label{cbmaalt}
c_{\beta-\alpha}=\frac{(\mathcal{M}_{11}^2-m_h^2)\cosb+\mathcal{M}_{12}^2\sinb}{\sqrt{(m_H^2-m_h^2)(\mathcal{M}_{11}^2-m_h^2)}}\,.
\ee
The exact alignment condition corresponds to the vanishing of the numerator in \eq{cbmaalt}, which yields
\be \label{condone}
t_\beta\mathcal{M}_{12}^2=m_h^2-\mathcal{M}^2_{11}\,.
\ee
Dividing \eq{condone} by $\mathcal{M}_{12}^4$ and using \eq{relation} then gives
\be \label{condtwo}
t_{\beta}^{-1}\mathcal{M}_{12}^2=m_h^2-\mathcal{M}^2_{22}\,.
\ee
Eliminating $m_h^2$ from \eqs{condone}{condtwo},
\be \label{exact}
c_{2\beta}\mathcal{M}_{12}^2=\sinb\cosb(\mathcal{M}_{11}^2-\mathcal{M}_{22}^2)\,.
\ee
Using \eqst{eq:cpemass}{eqL22}, one can check that \eq{exact} is equivalent to the condition
$Z_6=0$, where $Z_6$ is given by \eq{zeesix}. In addition, one can use either \eq{condone} or (\ref{condtwo}) to obtain $m_h^2=Z_1 v^2$, where $Z_1$ is given by \eq{zeeone}, as expected in light of \eq{alignmentconditions}.

In the 2HDM, the exact alignment limit of $Z_6=0$ can be achieved in four possible ways: (i) as a consequence of an exact symmetry of the theory;
(ii) as a consequence of an exact symmetry of the scalar potential, which is broken by the Higgs-Yukawa interactions; (iii)~as
a consequence of an accidental global symmetry of the scalar potential, which is broken by the gauge interactions and Higgs-fermion Yukawa interactions; or (iv)  
accidentally due to a choice of scalar potential parameters that is not governed by any symmetry.
We exhibit these four possibilities in turn.

An example of case (i) is the inert 2HDM~\cite{Barbieri:2006dq}.  In this model, 
the theory possesses an exact $\mathbb{Z}_2$ symmetry in the Higgs basis, under which the Higgs basis field $H_2$ is odd and all other fields ($H_1$, fermions and gauge bosons) are even.
In this case $Z_6=0$ as a consequence of the $\mathbb{Z}_2$ symmetry~\cite{whitepaper}, which remains unbroken in the vacuum since $\langle H_2^0\rangle=0$.

An example of case (ii) is the 2HDM with the scalar potential parameters of \eq{eq:generalpotential} given by~\cite{Ferreira:2009wh,Dev:2014yca} 
\be \label{sym}
m_{11}^2=m_{22}^2\,,\quad \lambda_1=\lambda_2=\lambda_3+\lambda_4+\lambda_5\,,\quad m_{12}^2=\lambda_6=\lambda_7=0\,.
\ee
These conditions on the $\lambda_i$ yield $Z_6=0$ [cf.~\eq{zeesix}].
\Eq{sym} is satisfied by 2HDM scalar potentials with a generalized CP3 symmetry or with an SO(3) Higgs flavor symmetry (the latter if $\lambda_5=0$ also holds), as shown in Ref.~\cite{Ferreira:2009wh}.
In general these two symmetries will not be respected by the Higgs-fermion Yukawa interactions~\cite{Ferreira:2010bm}.

Custodial symmetric
scalar potentials provide examples of case (iii).  Indeed, custodial symmetries~\cite{Sikivie:1980hm} are broken by the hypercharge gauge interactions as well as by the Higgs-fermion Yukawa
interactions.  The maximally symmetric 2HDM of Ref.~\cite{Dev:2014yca} with an SO(5) global symmetry, which yields \eq{sym} with $\lambda_4=\lambda_5=0$, provides an example of this case.
In particular, Ref.~\cite{Dev:2014yca} has stressed the role of the symmetries that lead to \eq{sym}, which yields exact alignment at tree-level.  Deviations
from alignment are generated due to loop effects, since these are not exact symmetries of the full theory.
 
Finally, as we shall see in the next subsection, \eq{sym} does not hold for the MSSM Higgs sector.  Thus, alignment 
can only arise for a special choice of parameters and is not a consequence of any symmetry.

For completeness, we record the Yukawa couplings of the two Higgs doublets to a single generation of up and down-type quarks.  Employing the notation of the third generation,
\be \label{thdmyuk}
-\mathscr{L}_{\rm Yuk}=\mathcal{Y}^1_b\overline{b}_R\Phi_1^{i\,*}Q^i_L
+\mathcal{Y}^2_b\overline{b}_R\Phi_2^{i\,*}Q^i_L
+\epsilon_{ij}\bigl[\mathcal{Y}^1_t\overline{t}_R Q_L^i\Phi^j_1+\mathcal{Y}^2_t\overline{t}_R Q_L^i\Phi^j_2\bigr]
+{\rm h.c.}\,,
\ee
where $\epsilon_{12}=-\epsilon_{21}=1$, $\epsilon_{11}=\epsilon_{22}=0$, $Q_L=(t_L\,,\,b_L)$ are the doublet left handed quark fields and $t_R$, $b_R$ are the singlet right-handed quark fields. Inserting $\langle\Phi_i^0\rangle=v_i/\sqrt{2}$ yields the quark masses,
\be \label{qmasses}
m_b=(v_1 \mathcal{Y}_b^1+v_2 \mathcal{Y}_b^2)/\sqrt{2}\,,\qquad\quad
m_t=(v_1 \mathcal{Y}_t^1+v_2 \mathcal{Y}_t^2)/\sqrt{2}\,.
\ee

\subsection{The MSSM Higgs Sector}

The Higgs sector of the MSSM is a 2HDM whose dimension-four couplings are constrained by supersymmetry. In particular, at tree-level,
\bea
\lambda_1&=&\lambda_2=-(\lambda_3+\lambda_4)=\tfrac{1}{4}(g^2+g^{\prime\,2})=m_Z^2/v^2\,,\label{mssmlam1}\\
\lambda_4&=&-\half g^2=-2m_W^2/v^2\,,\label{mssmlam2}\\
\lambda_5&=&\lambda_6=\lambda_7=0\,.\label{mssmlam3}
\eea
These results yield the well-known formulae for the tree-level MSSM CP-even Higgs masses.
At tree-level, $(m_h^2)\lsub{\rm max}=m_Z^2 c^2_{2\beta}$, which is not consistent with experimental data. However, radiative corrections can
have large contributions to the tree-level Higgs mass, and regions of MSSM parameter space can be found where $m_h\simeq 125$~GeV, as required by the data.

The mixing angle, which governs the Higgs couplings, is easily written down using the Higgs basis. Using \eqs{zeeone}{zeesix},\footnote{Note that $\beta$ has been promoted to a physical parameter, since the tree-level coupling relations given in \eqst{mssmlam1}{mssmlam3} are a consequence of supersymmetry, which establishes a preferred basis choice for the scalar Higgs fields.}
\be
Z_1 v^2=m_Z^2 c^2_{2\beta}\,,\qquad\quad
Z_6 v^2=-m_Z^2 s_{2\beta}c_{2\beta}\,.
\ee
Inserting the above results into \eq{cbmaf} yields the tree-level result,
\be
c_{\beta-\alpha}=\frac{m_Z^2 \,s_{2\beta}c_{2\beta}}{\sqrt{(m_H^2-m_h^2)(m_H^2-m_Z^2 c^2_{2\beta})}}\,.
\ee
In the decoupling limit, one recovers \eq{dl} as expected.  In addition, radiative corrections
that are required to yield a phenomenologically acceptable value of $m_h$, do not significantly modify the decoupling behavior exhibited above. In contrast, alignment cannot be achieved without decoupling at tree-level (except at the endpoints where either $s_\beta = 0$ or $c_\beta = 0$, for which no tree-level mass is obtained for the up-type and down-type quarks, respectively, and at the midpoint $t_\beta = 1$, which leads to a vanishing lightest CP-even Higgs mass at tree-level.  None of these scenarios are experimentally viable.). We shall see in the next subsection that including radiative corrections,  alignment  independent of decoupling can be achieved in the MSSM at values of $\beta$ away from the endpoints, resulting in important phenomenological consequences.

Supersymmetry also imposes constraints on the Higgs-fermion interactions. In the supersymmetric literature, it is common to define:
\be \label{HUHD}
H_D^i\equiv \epsilon_{ij}\Phi_1^{j\,*}\,,\qquad\quad H_U^i=\Phi_2^i\,.
\ee In terms of $H_U$ and $H_D$, the Yukawa couplings given in \eq{thdmyuk} must be holomorphic, which implies that $\mathcal{Y}^1_t=\mathcal{Y}^2_b=0$.  This yields the so-called Type-II Higgs--quark couplings,\footnote{As in the previous subsection, we neglect the full generation structure of the Yukawa couplings and focus on the couplings of the Higgs bosons to the third generation quarks.}
\be
-\mathscr{L}_{\rm Yuk}=\epsilon_{ij}\bigl[h_b \overline{b}_R H_D^i Q^j_L+h_t\overline{t}_R Q^i_L H_U^j\bigr]+{\rm h.c.}\,,
\ee
where we have resorted to the more common notation $h_b=\mathcal{Y}_b^1$ and $h_t=\mathcal{Y}_t^2$.  \Eq{qmasses} then yields:
\be \label{tbmasses}
m_b= h_b v c_\beta/\sqrt{2}\,,\qquad\quad  m_t=h_t v s_\beta/\sqrt{2}\,.
\ee
The corresponding tree-level Yukawa couplings of the lightest CP-even Higgs boson to
down-type and up-type quark pairs are given by
\bea
g_{h b\bar b} &= &-\frac{m_b}{v}\,\frac{s_\alpha}{c_\beta}= \frac{m_b}{v}\,\bigl(\sbma-\cbma t_\beta\bigr)\,,
\label{hlbbtree}  \\[5pt]
g_{h t\bar t} & = & \phm\frac{m_t}{v}\,\frac{c_\alpha}{s_\beta}=\frac{m_t}{v}\,\bigl(\sbma+\cbma t_\beta^{-1}\bigr)\,.
\label{hltttree}
\eea
\Eqs{hlbbtree}{hltttree} exhibit the expected behavior in the decoupling/alignment limits.  That is, when $c_{\beta-\alpha}=0$, we recover the SM result, $g_{hf\bar{f}}=m_f/v$. However, note that in the absence of exact alignment, the deviation from SM couplings of the down-type Yukawa coupling, is $t_\beta$ enhanced. Therefore, it is not enough to demand $|c_{\beta-\alpha}| \ll1$. Rather, proper SM-like behavior of the coupling of $h$ to down-type quarks is recovered  if $|c_{\beta-\alpha}| \ll 1/t_\beta$.  This phenomenon has been called \textit{delayed decoupling} in Refs.~\cite{Haber:2000kq,Gunion:2002zf,CHLM,Ferreira:2014naa}.

In the MSSM, the coupling of the Higgs bosons to squarks and sleptons are governed by both supersymmetry-conserving and supersymmetry-breaking parameters.   The relevant couplings can be found in Ref.~\cite{Gunion:1984yn}.  For later use, we shall focus here on the couplings of $H_U$ and $H_D$ to the third generation squarks that are proportional to the Higgs--top quark Yukawa coupling, $h_t$.  The corresponding terms in the interaction Lagrangian are
\be \label{squarks}
\mathscr{L}_{\rm int}\supset h_t\bigl[\mu^*(H_D^\dagger \widetilde Q) \widetilde U + A_t\epsilon_{ij}
H_U^i\widetilde Q^j\widetilde U+{\rm h.c.}\bigr]-h_t^2\bigl[H_U^\dagger H_U(\widetilde Q^\dagger\widetilde Q+\widetilde U^*\widetilde U)-|\widetilde Q^\dagger H_U|^2\bigr]\,,
\ee
with an implicit sum over the weak $SU(2)$ indices $i,j=1,2$, where in the notation of Ref.~\cite{Gunion:1984yn},
\be
\widetilde Q=\begin{pmatrix} \widetilde t_L\\ \widetilde b_L\end{pmatrix}\,,\qquad\quad
\widetilde U\equiv \widetilde t_R^*\,,
\ee
and in general the supersymmetric Higgsino mass parameter, $\mu$, and the supersymmetry-breaking parameter, $A_t$, are complex.

It is convenient to rewrite \eq{squarks} in terms of the Higgs basis fields.  Using \eqs{cphiggsbasisfields}{HUHD}, it follows that
\bea \label{squarkshbasis}
\mathscr{L}_{\rm int}&\supset &h_t\epsilon_{ij}\bigl[(\sinb X_t H_1^i+\cosb Y_t
H_2^i)\widetilde Q^j\widetilde U+{\rm h.c.}\bigr]\nn\\[6pt]
&&\quad -h_t^2\biggl\{\biggl[s_\beta^2|H_1|^2+c_\beta^2|H_2|^2
+\sinb\cosb(H_1^\dagger H_2+{\rm h.c.})\biggr](\widetilde Q^\dagger\widetilde Q+\widetilde U^*\widetilde U)\nn\\
&&\qquad\qquad -s_{\beta}^2|\widetilde Q^\dagger H_1|^2-c_\beta^2|\widetilde Q^\dagger H_2|^2-
\sinb\cosb\bigl[(\widetilde Q^\dagger H_1)(H_2^\dagger\widetilde Q)+{\rm h.c.}\bigr]\biggr\}\,,
\eea
where
\be \label{XY}
X_t\equiv A_t-\mu^*/t_\beta\,,\qquad\quad Y_t\equiv A_t+\mu^*t_\beta\,.
\ee
Note that the terms proportional to $X_t$ in \eq{squarkshbasis} are responsible for the mixing of $\widetilde t_L$ and $\widetilde t_R$ in the top-squark squared-mass matrix; the corresponding off-diagonal element is $(\mathcal{M}^2_{\tilde t})_{LR}= m_t X_t$, after setting $\langle H_1^0\rangle=v/\sqrt{2}$ and using \eq{tbmasses}. For simplicity, we shall henceforth assume that $\mu$ and $A_t$ are real, thereby neglecting possible CP-violating effects that can be introduced into the MSSM Higgs sector via radiative corrections.

Radiative corrections play a critical role in the MSSM Higgs sector. Three important mass scales are relevant---the scale of the squark masses, denoted by $M_S$, the mass of $h$ or $Z$ (which represents the electroweak scale) and the mass scale of the non-standard Higgs bosons, $H$, $A$ and $H^\pm$, which we will usually take to be $m_A$.  We shall assume that $M_S\gg m_A$.   In this case, we can formally integrate out the squarks to obtain a low-energy effective theory below the scale $M_S$, which is a general 2HDM with quartic and fermion couplings determined by their Type-II tree-level values plus radiative corrections induced by supersymmetry breaking effects.   Since the lightest CP-even Higgs boson couplings have been measured to be close to the the SM values, we infer that either we are in the decoupling limit, $m_h\ll m_A\ll M_S$ ,or the alignment limit independent of decoupling, $m_h\lsim m_A\ll M_S$.   In practice, the alignment limit independent of decoupling is most relevant for $m_A\,,\,m_H< 2m_t$.  For heavier values of $m_A$, the behavior of the Higgs sector approaches that of  the decoupling regime.

After integrating out the squarks, the supersymmetric relations that govern the scalar potential parameters [given in \eqst{mssmlam1}{mssmlam3}] are modified.  At one loop, the leading logarithmic corrections, which only appear for $\lambda_1\,\ldots,\lambda_4$, can be found in \Ref{Haber:1993an}. In addition, threshold corrections proportional to the MSSM parameters, $A_t$, $A_b$ and $\mu$, can also contribute significant corrections to all the scalar potential parameters, $\lambda_1\,\ldots,\lambda_7$. The relevant expressions are rather lengthy.  To get a sense of the corrections, we note that the largest contributions are proportional to the fourth power of the top-quark Yukawa coupling, $h_t$. Using the results given in \Ref{Haber:1993an} (the corresponding leading two-loop corrections to the quartic couplings can be found in Ref.~\cite{CEQW}), we obtain the following expressions for $Z_1$, $Z_5$ and $Z_6$ [cf.~\eqst{zeeone}{zeesix}] in the limit of $m_Z$, $m_A\ll M_S$, which include all one-loop radiative corrections proportional to $h_t^4$,
\bea
Z_1 v^2 &=& m_Z^2 c^2_{2\beta}+\frac{3v^2 s_\beta^4 h_t^4}{8\pi^2}\left[\ln\left(\frac{M_S^2}{m_t^2}\right)+\frac{X_t^2}{M_S^2}\left(1-\frac{X_t^2}{12M_S^2}\right)\right]\,,\label{mhmax}\\[8pt]
Z_5 v^2  & = & s_{2\beta}^2\left\{m_Z^2+\frac{3v^2 h_t^4}{32\pi^2}\left[\ln\left(\frac{m_S^2}{m_t^2}\right)+\frac{X_t Y_t}{m_S^2}\left(1-\frac{X_t Y_t}{12m_S^2}\right)\right]\right\}\,,
\label{mHcorr}\\[8pt]
Z_6 v^2&=& -s_{2\beta}\left\{m_Z^2 c_{2\beta}-\frac{3v^2 s_\beta^2  h_t^4}{16\pi^2}\biggl[\ln\left(\frac{M_S^2}{m_t^2}\right)+\frac{X_t(X_t+Y_t)}{2M_S^2}-\frac{X_t^3 Y_t}{12 M_S^4}\biggr]\right\},
\label{zeesixcorr} 
\eea
where $X_t$ and $Y_t$ are given by \eq{XY}. The upper bound for the squared-mass of the lightest CP-even Higgs boson is given by $(m_h^2)_{\rm max}= Z_1 v^2$. Indeed, \eq{mhmax} exhibits the well-known leading one-loop approximation for the upper bound on $m_h^2$ in the MSSM.

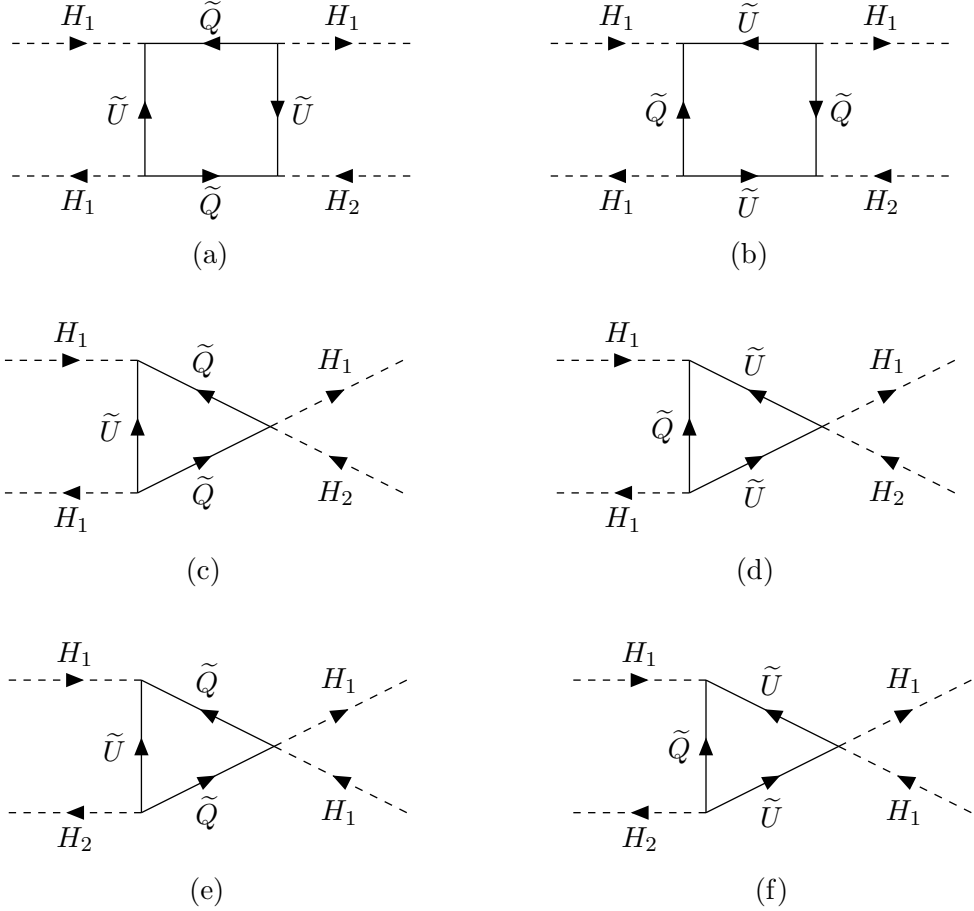
\begin{figure}[t]
\begin{center}
\begin{picture}(100,100)(-40,0)
\DashArrowLine(50,0)(0,0){3}
\DashArrowLine(0,50)(50,50){3}
\ArrowLine(50,0)(50,50)
\ArrowLine(50,0)(100,0)
\ArrowLine(100,50)(100,0)
\ArrowLine(100,50)(50,50)
\DashArrowLine(150,0)(100,0){3}
\DashArrowLine(100,50)(150,50){3}
\Text(25,60)[]{$H_1$}
\Text(25,-10)[]{$H_1$}
\Text(75,60)[]{$\widetilde Q$}
\Text(75,-10)[]{$\widetilde Q$}
\Text(40,25)[]{$\widetilde U$}
\Text(110,25)[]{$\widetilde U$}
\Text(125,60)[]{$H_1$}
\Text(125,-10)[]{$H_2$}
\Text(75,-30)[]{(a)}
\end{picture}
\begin{picture}(100,100)(-140,0)
\DashArrowLine(50,0)(0,0){3}
\DashArrowLine(0,50)(50,50){3}
\ArrowLine(50,0)(50,50)
\ArrowLine(50,0)(100,0)
\ArrowLine(100,50)(100,0)
\ArrowLine(100,50)(50,50)
\DashArrowLine(150,0)(100,0){3}
\DashArrowLine(100,50)(150,50){3}
\Text(25,60)[]{$H_1$}
\Text(25,-10)[]{$H_1$}
\Text(75,60)[]{$\widetilde U$}
\Text(75,-10)[]{$\widetilde U$}
\Text(40,25)[]{$\widetilde Q$}
\Text(110,25)[]{$\widetilde Q$}
\Text(125,60)[]{$H_1$}
\Text(125,-10)[]{$H_2$}
\Text(75,-30)[]{(b)}
\end{picture}
\begin{picture}(100,100)(170,120)
\DashArrowLine(50,0)(0,0){3}
\DashArrowLine(0,50)(50,50){3}
\ArrowLine(50,0)(50,50)
\ArrowLine(50,0)(100,25)
\ArrowLine(100,25)(50,50)
\DashArrowLine(150,0)(100,25){3}
\DashArrowLine(100,25)(150,50){3}
\Text(25,60)[]{$H_1$}
\Text(25,-10)[]{$H_1$}
\Text(75,50)[]{$\widetilde Q$}
\Text(75,0)[]{$\widetilde Q$}
\Text(40,25)[]{$\widetilde U$}
\Text(125,50)[]{$H_1$}
\Text(125,0)[]{$H_2$}
\Text(75,-30)[]{(c)}
\end{picture}
\begin{picture}(100,100)(65,120)
\DashArrowLine(50,0)(0,0){3}
\DashArrowLine(0,50)(50,50){3}
\ArrowLine(50,0)(50,50)
\ArrowLine(50,0)(100,25)
\ArrowLine(100,25)(50,50)
\DashArrowLine(150,0)(100,25){3}
\DashArrowLine(100,25)(150,50){3}
\Text(25,60)[]{$H_1$}
\Text(25,-10)[]{$H_1$}
\Text(75,50)[]{$\widetilde U$}
\Text(75,0)[]{$\widetilde U$}
\Text(40,25)[]{$\widetilde Q$}
\Text(125,50)[]{$H_1$}
\Text(125,0)[]{$H_2$}
\Text(75,-30)[]{(d)}
\end{picture}
\begin{picture}(100,100)(65,140)
\DashArrowLine(50,0)(0,0){3}
\DashArrowLine(0,50)(50,50){3}
\ArrowLine(50,0)(50,50)
\ArrowLine(50,0)(100,25)
\ArrowLine(100,25)(50,50)
\DashArrowLine(150,0)(100,25){3}
\DashArrowLine(100,25)(150,50){3}
\Text(25,60)[]{$H_1$}
\Text(25,-10)[]{$H_2$}
\Text(75,50)[]{$\widetilde Q$}
\Text(75,0)[]{$\widetilde Q$}
\Text(40,25)[]{$\widetilde U$}
\Text(125,50)[]{$H_1$}
\Text(125,0)[]{$H_1$}
\Text(75,-30)[]{(e)}
\end{picture}
\begin{picture}(100,100)(-45,140)
\DashArrowLine(50,0)(0,0){3}
\DashArrowLine(0,50)(50,50){3}
\ArrowLine(50,0)(50,50)
\ArrowLine(50,0)(100,25)
\ArrowLine(100,25)(50,50)
\DashArrowLine(150,0)(100,25){3}
\DashArrowLine(100,25)(150,50){3}
\Text(25,60)[]{$H_1$}
\Text(25,-10)[]{$H_2$}
\Text(75,50)[]{$\widetilde U$}
\Text(75,0)[]{$\widetilde U$}
\Text(40,25)[]{$\widetilde Q$}
\Text(125,50)[]{$H_1$}
\Text(125,0)[]{$H_1$}
\Text(75,-30)[]{(f)}
\end{picture}
\end{center}
\vskip 2.3in
\caption{\label{thresholdcorrs}{\em One-loop diagrams contributing to the
the coefficient, $Z_6$, of the Higgs basis operator, $(H_1^\dagger H_1)(H_1^\dagger H_2)$.
Using the interaction Lagrangian given in \eq{squarkshbasis}, one sees that the parametric
dependence for the six diagrams are: $h_t^4 s_\beta^3 c_\beta X_t^3 Y_t$ for (a) and (b);
$h_t^4 s_\beta^3 c_\beta X_t^2$ for (c) and (d); and $h_t^4 s_\beta^3 c_\beta X_t Y_t$ for (e) and (f).
}}
\end{figure}

The structure of the threshold corrections [proportional to either $X_t$ or $Y_t$ in \eqthree{mhmax}{zeesixcorr}{mHcorr}] is easy to understand.  For example, in Fig.~\ref{thresholdcorrs}, we exhibit the leading one-loop corrections to $Z_6$, which corresponds to the coefficient of the operator $[(H_1^\dagger H_1)(H_1^\dagger H_2)+{\rm h.c.}]$ [cf.~\eq{higgsbasispot}] in the Higgs basis.  Using the interaction Lagrangian given  by \eq{squarkshbasis}, one can immediately ascertain the parametric dependence of the diagrams shown in Fig.~\ref{thresholdcorrs}.  Each diagram has a $s_\beta^3\cosb h_t^4$ dependence, and there is a factor of $X_t$ [$Y_t$] for each  $H_1\widetilde Q\widetilde U$~[$H_2\widetilde Q\widetilde U$] vertex, respectively.  In this way, we explain the parametric dependence of the threshold corrections to $Z_6$ exhibited in \eq{zeesixcorr}.  Likewise, by replacing the external $H_2$ [$H_1$] line with an $H_1$ [$H_2$] line in Fig.~\ref{thresholdcorrs} and deleting graphs (e) and (f), which are now identical to graphs (c) and (d), we can understand the parametric dependence of the threshold corrections to $Z_1$ [$Z_5$].

It is instructive to obtain an approximate one-loop formula for $c_{\beta-\alpha}$, keeping only the leading $\mathcal{O}(h_t^4)$ corrections.  We can also simplify the result by considering the large
$t_\beta$ limit.  Indeed, the resulting expressions will provide good approximations for $t_\beta\gsim 5$ (a region of considerable interest in our analysis).  In the large $t_\beta$ limit, we may
approximate $s_\beta\simeq 1$ and $c_{2\beta}\simeq -1$.  Moreover, in this approximation the radiatively corrected value of the squared-mass of the light CP-even Higgs boson at one-loop is
\be \label{mhtwomax}
m_h^2\simeq Z_1 v^2\simeq m_Z^2+\frac{3m_t^4}{2\pi^2 v^2}\left[\ln\left(\frac{M_S^2}{m_t^2}\right)+\frac{X_t^2}{M_S^2}\left(1-\frac{X_t^2}{12M_S^2}\right)\right]\,,
\ee
where we have used \eq{tbmasses} to write $v^2\sinb^4 h_t^4=4m_t^4/v^2$. Using \eqs{zeesixcorr}{mhtwomax} in the evaluation of \eq{cbmaf} yields
\be \label{tbcbma}
t_\beta\; c_{\beta-\alpha}\simeq \frac{-1}{m_H^2-m_h^2}\left[m_h^2+m_Z^2+
\frac{3m_t^4 X_t(Y_t-X_t)}{4\pi^2 v^2
  M_S^2}\left(1-\frac{X_t^2}{6M_S^2}\right)\right]\,.
\ee
At large $t_\beta$ we have $X_t(Y_t-X_t)\simeq \mu(A_t t_\beta-\mu)$ and $X^3_t(Y_t-X_t)\simeq \mu A_t^2(A_tt_\beta-3\mu)$, in which case, \eq{tbcbma} can be rewritten in the following approximate form,
\be \label{tanbcbma}
t_\beta\; c_{\beta-\alpha}\simeq \frac{-1}{m_H^2-m_h^2}\left[m_h^2+m_Z^2+
\frac{3m_t^4 }{4\pi^2 v^2
  M_S^2}\left\{A_t\mu t_\beta\left(1-\frac{A_t^2}{6M_S^2}\right)-\mu^2\left(1-\frac{A_t^2}{2M_S^2}\right)\right\}\right]\,.
\ee
The significance of the product $t_\beta\; c_{\beta-\alpha}$ has already been noted below
\eq{hltttree}.  Namely, the condition that guarantees that the coupling of $h$ to down-type quarks and leptons is close to its SM value is $t_\beta |c_{\beta-\alpha}|\ll 1$.  In contrast, all other $h$ couplings approach their SM values for $|c_{\beta-\alpha}|\ll 1$, independently of the value of $t_\beta$.

The Higgs-fermion Yukawa couplings are also modified below the scale $M_S$.  Having integrated out the squarks, the low-energy effective Yukawa couplings are no longer of Type-II (which had been previously enforced by supersymmetry). The Yukawa couplings below the scale $M_S$ have the form given in \eq{thdmyuk},
\be \label{thdmsusyyuk}
-\mathscr{L}_{\rm Yuk}=\epsilon_{ij}\bigl[(h_b+\delta h_b)\overline{b}_R H_D^{i}Q^j_L
+(h_t+\delta h_t)\overline{t}_R Q_L^i H^j_U\bigr]
+\Delta h_b \overline{b}_R Q^i_L H_U^{i\,*}+\Delta h_t \overline{t}_R Q_L^i H^{i\,*}_D
+{\rm h.c.}\,,
\ee
where $\delta h_{t,b}$ and $\Delta h_{t,b}$ represent one-loop corrections from squark/gaugino loops. \Eq{thdmsusyyuk} yields a modification of the tree-level relations between $h_t$, $h_b$ and $m_t$, $m_b$ as follows~\cite{deltamb}:
\bea
        m_b &=& \frac{h_b v}{\sqrt{2}} c_\beta
        \left(1 + \frac{\delta h_b}{h_b}
        + \frac{\Delta h_b t_\beta}{h_b} \right)
        \equiv\frac{h_b v}{\sqrt{2}} c_\beta
        (1 + \Delta_b)\,, \label{byukmassrel}\\
        &&\nonumber\\
        m_t &=& \frac{h_t v}{\sqrt{2}} s_\beta
        \left(1 + \frac{\delta h_t}{h_t} + \frac{\Delta
        h_t\cot\beta}{h_t} \right)
        \equiv\frac{h_t v}{\sqrt{2}} s_\beta
        (1 + \Delta_t)\,, \label{tyukmassrel}
\eea
which define the quantities $\Delta_b$ and $\Delta_t$.\footnote{The dominant contributions to $\Delta_b$ are $t_\beta$-enhanced, with $\Delta_b\simeq (\Delta h_b/h_b)t_\beta$; for $t_\beta\gg 1$, $\delta h_b/h_b$ provides a small correction to $\Delta_b$. In the same limit, $\Delta_t\simeq\delta h_t/h_t$, with the additional contribution of $(\Delta h_t/h_t)\cot\beta$ providing a small correction.  In practical applications, it is often sufficient to keep only $\Delta_b$, which provides the dominant contributions to the radiatively-corrected Yukawa couplings.} Diagonalizing the CP-even Higgs squared-mass matrix, \eqst{thdmsusyyuk}{tyukmassrel} then yield the physical couplings of $h$ to the up-type and down-type quarks.  After resummation of the dominant corrections~\cite{Carena:1999py,CMW,mssmhiggsreview1}, the resulting expressions can be written in the following forms:
\bea
g_{h b\bar b} &= & \frac{m_b}{v}\left[\sbma-\cbma t_\beta
-\frac{1}{1+\Delta_b}\left(\frac{\delta h_b}{h_b}-
\Delta_b\right)\left(\frac{\cbma t_\beta}{s^2_\beta}\right)\right]\,,
\label{hlbb}  \\[5pt]
g_{h t\bar t} & = & \frac{m_t}{v}\left[\sbma+\cbma t_\beta^{-1}
-\frac{1}{1+\Delta_t}\frac{\Delta h_t}{h_t}
\left(\frac{\cbma}{s^2_\beta}\right)\right]\,.
\label{hltt}
\eea
Note that the radiative corrections to the couplings of $h$ to the up-type and down-type quarks
vanish in the limit of exact alignment where $c_{\beta-\alpha}=0$.  However, the phenomenon of delayed decoupling at large $t_\beta$, discussed below \eq{hltttree}, persists.  That is,
at large values of $t_\beta$, the $hb\bar{b}$ coupling approaches the corresponding SM value in the limit of $t_\beta|c_{\beta-\alpha}|\ll 1$.

\subsection{Alignment Independent of Decoupling in the MSSM Higgs Sector}

In the previous section, we noted that alignment independent of decoupling is not possible for the tree-level MSSM Higgs sector, since $Z_6 v^2=-m_Z^2 s_{2\beta} c_{2\beta}\neq 0$, except at phenomenologically unacceptable values of $\beta$.   Once radiative corrections are included, alignment independent of decoupling can occur quite generically, due to the appearance of a branch of solutions that are absent at tree level \cite{Carena:2013ooa}.

To exhibit explicitly the cancellation that yields alignment, we make use of the fact that exact alignment is attained when $Z_6=0$.   Assuming that $s_{2\beta}\neq 0$, it then follows from \eq{zeesixcorr} that exact alignment at one-loop order is achieved when
\be \label{zeesixzero}
m_Z^2 c_{2\beta}=\frac{3v^2 s_\beta^2  h_t^4}{16\pi^2}\biggl[\ln\left(\frac{M_S^2}{m_t^2}\right)+\frac{X_t(X_t+Y_t)}{2M_S^2}-\frac{X_t^3 Y_t}{12 M_S^4}\biggr]\,,
\ee
where $X_t$ and $Y_t$ are defined in \eq{XY}.  \Eq{zeesixzero} yields a non-linear polynomial equation for $t_\beta$.  If a solution exists for positive $t_\beta$ (recall that $0\leq\beta\leq\half\pi$ by convention) for fixed values of the other MSSM parameters, then the alignment limit can be realized.
To exhibit that a solution is possible, we shall assume that $t_\beta\gg 1$ (in practice, moderate to large values of $t_{\beta}\gsim 5$ are sufficient).   We then perform a Taylor expansion of \eq{zeesixzero} keeping only constant terms and terms linear in $t_{\beta}^{-1}$.  We can then easily solve for $t_\beta$,
\be \label{tbsol}
t_\beta=\frac{m_Z^2+\displaystyle\frac{3 v^2 h_t^4}{16\pi^2}\left[\ln\left(\frac{M_S^2}{m_t^2}\right)
+\displaystyle\frac{2A_t^2-\mu^2}{2M_S^2}-\displaystyle\frac{A_t^2(A_t^2-3\mu^2)}{12M_S^4}\right]}
{\displaystyle\frac{3v^2 h_t^4 \mu A_t}{32\pi^2 M_S^2}\left(\displaystyle\frac{A_t^2}{6M_S^2}-1\right)}\,.
\ee

Since we have assumed that $t_\beta\gg 1$ in deriving \eq{tbsol}, we can rewrite this result in terms of $m_h^2$ [cf.~\eq{mhtwomax}] and $m^4_t$ (after taking $\sinb\simeq 1$),\footnote{As a check of \eq{tbsolalt}, one can verify that the same result is obtained by setting the approximate expression of $c_{\beta-\alpha}$ obtained in \eq{tanbcbma} to zero.}
\be \label{tbsolalt}
t_\beta=\frac{m_h^2+m_Z^2+\displaystyle\frac{3m_t^4\mu^2}{4\pi^2 v^2M_S^2}\left(\frac{A_t^2}{2M_S^2}-1\right)}
{\displaystyle\frac{3m_t^4 \mu A_t}{4\pi^2 v^2 M_S^2}\left(\displaystyle\frac{A_t^2}{6M_S^2}-1\right)}\,.
\ee
For values of $\mu$, $A_t\sim\mathcal{O}(M_S)$, the term of $\mathcal{O}(m_t^4)$ in the numerator of \eq{tbsolalt} is subdominant. Since $t_\beta$ is positive, it follows that a viable solution exists if $\mu A_t(A_t-\sqrt{6}M_S)>0$. In the approximations employed in obtaining \eq{mhtwomax}, the so-called maximal mixing condition, that yields the largest radiatively-corrected Higgs mass, corresponds to $A_t=\sqrt{6}\,M_S$.  Moreover, one obtains $t_{\beta}\gg 1$ if $\mu A_t >0$ [$\mu A_t<0$] with values of $A_t$ not too far above [below] the maximal mixing condition, which is consistent with the assumption used in the derivation of  \eq{tbsolalt}.

To make contact again with the results of Ref.~\cite{Carena:2013ooa}, we observe that the exact alignment condition, $Z_6=0$,  is achieved when [cf.~\eq{zeesix}]:
\be
(\lambda_1-\lambda_{345})\cosbii-(\lambda_2-\lambda_{345})\sinbii=
(\cosbii-3\sinbii) t_\beta^{-1}\lambda_6+(3\cosbii-\sinbii)t_\beta\lambda_7\,,
\ee
where $\lambda_{345}\equiv(\lambda_3+\lambda_4+\lambda_5)$.
For $t_\beta\gg 1$, we can approximate $\cosb\sim t_\beta^{-1}\simeq 0$ and $\sinb\simeq 1$.  We then obtain Eq.~(103) of Ref.~\cite{Carena:2013ooa},
\be \label{tsol}
t_\beta\simeq \frac{\lambda_2-\lambda_{345}}{\lambda_7}\,.
\ee
The value of $t_\beta$ at which alignment takes place is inversely proportional to $\lambda_7$, which vanishes in the MSSM at  tree-level and arises only radiatively.\footnote{Using the radiatively corrected expressions for the couplings in \eq{tsol} given in \Ref{Haber:1993an}, keeping only terms proportional to $h_t^4$, we recover the expression given in \eq{tbsol}.}
As can be seen from Eq.~(\ref{tsol}), alignment at smaller $t_\beta$ requires a larger $\lambda_7$, unless there is a tuning between $\lambda_2$ and $\lambda_{345}$ in the numerator. In the end, it was found in  Ref.~\cite{Carena:2013ooa}, that for generic choices of parameters in the MSSM, alignment independent of decoupling typically occurs at some value of $t_\beta\agt 10$, with smaller $t_\beta$ requiring larger values of $A_t/M_S$ and $\mu/M_S$~[cf.~Eq.~(\ref{tbsol}].

For top squark masses of the order of a few TeV, the requirement of obtaining the proper value of $m_h$ constrains the values of $A_t \simlt 3 M_S$.  In Ref.~\cite{Carena:2013ooa} it was demonstrated that alignment independent of decoupling may be obtained for $t_\beta$ of order 10 for large values of $\mu \simgt 2 M_S$ and for either positive values of $A_t$ of about $3 M_S$ or negative values of $A_t \simeq -1.5 M_S$.  Alignment values of $t_\beta < 10$ are not easily realized in the MSSM.\footnote{Alignment independent of decoupling for smaller values of $t_\beta$ may be obtained in the NMSSM~\cite{Carena:2013ooa} or in triplet extensions of the MSSM~\cite{Delgado:2013zfa}.}

\section{Searches for Heavy Higgs Bosons}

Our purpose is to study the interplay of direct searches and precision Higgs measurements in scenarios where alignment occurs at very large versus moderate $t_\beta$. In order to analyze the bounds on the non-standard Higgs masses, we choose  benchmark scenarios close to the ones proposed in Ref.~\cite{Carena:2013qia}, which are used by the LHC experimental collaborations in their analyses of searches for non-standard Higgs bosons (see, for example, Refs.~\cite{Khachatryan:2014wca,Aad:2014vgg}). Specifically, in Table \ref{tab:1} we define two classes of benchmarks, $m_{h}^{\rm mod+}$ and $m_{h}^{\rm alt}$, where the main difference with the $m_h^{\rm mod+}$  and the tau-phobic scenarios defined in Ref.~\cite{Carena:2013qia} is that we take $\mu$ and $m_Q$ as floating parameters.

\begin{table}[tb]
\begin{tabular}{ | c | c | c  | }
\hline
 & $m_{h}^{\rm alt }$ & $m_{h}^{\rm mod+} $ \\
\hline
$A_t/m_Q$ & 2.45 &1.5   \\
$M_2 = 2 \ M_1$ & 200 GeV & 200 GeV \\
$M_3$                 & 1.5 TeV   & 1.5 TeV \\
$m_{\tilde \ell} = m_{\tilde q}$   &  $m_Q$  & $m_Q$ \\
$A_\ell$ = $A_q$ &   $A_t$  & $A_t$ \\
\hline
\end{tabular}
\caption{\label{tab:1}\em Parameters in the on-shell scheme defining the $m_{h}^{\rm mod+}$ and $m_{h}^{\rm alt}$ scenarios. We leave $m_Q$ and $\mu$ as floating parameters.}
\end{table}

These two classes of scenarios differ in the choice of the ratio $A_t/m_Q$, which results in no alignment or alignment at very large values of $t_\beta$ for $m_h^{\rm mod+}$ and alignment at $t_\beta\alt 50$ for $m_h^{\rm alt}$ \cite{Carena:2013ooa}.  Although these benchmarks are inspired by those proposed in Ref.~\cite{Carena:2013qia}, the fact that we allow the $\mu$ parameter and the overall soft scale, $m_Q$, to vary allows us to obtain the correct mass for the lightest CP-even Higgs boson at small $t_\beta\alt 6$, and to study the impact of alignment at different values of  $t_\beta$.  Both have a crucial impact on the properties of the lightest CP-even Higgs boson and on the decays of the heavy CP-even and CP-odd Higgs bosons. Observe also that we fix the value of $A_t$ instead of $X_t$, as was done in Ref.~\cite{Carena:2013qia}, which makes a difference only at large values of $\mu$  and small values of $t_\beta \simlt 10$.  In particular our m$_h^{\rm mod+}$ scenario with $\mu = 200$~GeV has the same properties
as the m$_h^{\rm mod+}$ scenario in Ref.~\cite{Carena:2013qia} and we have therefore adapted the notation from that reference. All of our numerical results are obtained from {\tt FeynHiggs}~\cite{FeynHiggs}, which allows for a computation of all the relevant production cross sections and branching ratios.\footnote{It should be noted that there are relevant difference between the results obtained by {\tt FeynHiggs} and other higher order computations~\cite{Draper:2013oza,Bagnaschi:2014rsa,Degrassi:2014pfa}, but the analysis of the origin of these differences is beyond the scope of this article.}

Before discussing the details of the Higgs phenomenology, recall the approximate analytical expressions given in the previous section governing the behavior of the various couplings, for example, $c_{\beta-\alpha}$ obtained in \eq{tanbcbma}. In our benchmark scenarios, $m_Q$ denotes the common squark/slepton mass, hence one can identify $M_S=m_Q$. It should be noted that \eq{tanbcbma} does not include two-loop corrections, which can be significant. These two loop corrections approximately preserve the parametric dependence of our analytic expressions on $A_t/m_Q$ in the $\overline{\rm MS}$ and $\overline{\rm DR}$ schemes. This is not true in the on-shell scheme, which is employed in {\tt FeynHiggs}. Therefore, in comparing our analytic expressions with our numerical results, one should use the values of $A_t/m_Q$ in the $\overline{\rm MS}$ or $\overline{\rm DR}$ schemes, that are approximately $20\%$ larger than the ones in the on-shell scheme~\cite{mssmhiggsupperbound}.

\subsection{Getting the Correct $\boldsymbol{m_h}$ Everywhere}
\label{sect:higgsmass}

In scenarios defined previously in Ref.~\cite{Carena:2013qia}, stop masses are fixed at the order of  1 TeV, which fails to reproduce the proper lightest CP-even Higgs mass, $m_h \simeq 125$~GeV, at values of $t_\beta \leq 6$ (the precise value of $t_\beta$ at which this occurs depends on the specific scenario). In our benchmarks we vary the overall stop mass scale, $m_Q$, so that the lightest CP-even Higgs mass is in the experimentally observed range within theoretical uncertainties, which we take to be of the order of 3 GeV, $m_h = 125\pm3$ GeV. More specifically, for a given value of $t_\beta$, $\mu/m_Q$ and $A_t/m_Q$, we fix the value of $m_Q$  for small values of $m_A \simeq 200$~GeV in such a way that the lightest CP-even Higgs mass is about 123~GeV.  This is enough to keep the value of $m_h$ in the acceptable range for all values of $m_A$.\footnote{In the $m_h^{\rm alt}$ scenario for $\mu=3m_Q$, $m_A\sim 200$ GeV and $t_\beta \gtrsim 40$, the Higgs mass is somewhat lower than 123~GeV due to sbottom effects. However, this region of parameter space is excluded regardless of the light Higgs mass, therefore we do not tune the value of $m_Q$ in this region.} The small variation of the lightest CP-even Higgs mass for larger values of $m_A$ has only a minor impact on the heavy Higgs phenomenology and does not affect the signal strength of the lightest CP-even Higgs in any significant way.   In contrast, fixing the value of $m_Q$ at around 1 TeV, as currently done by the experimental collaborations, leads to artificially low values of $m_h$ at low values of $t_\beta$ that can have a large impact on the Higgs boson phenomenology.

\begin{figure}[ht]
\subfloat[]{\includegraphics[width=3.2in, angle=0]{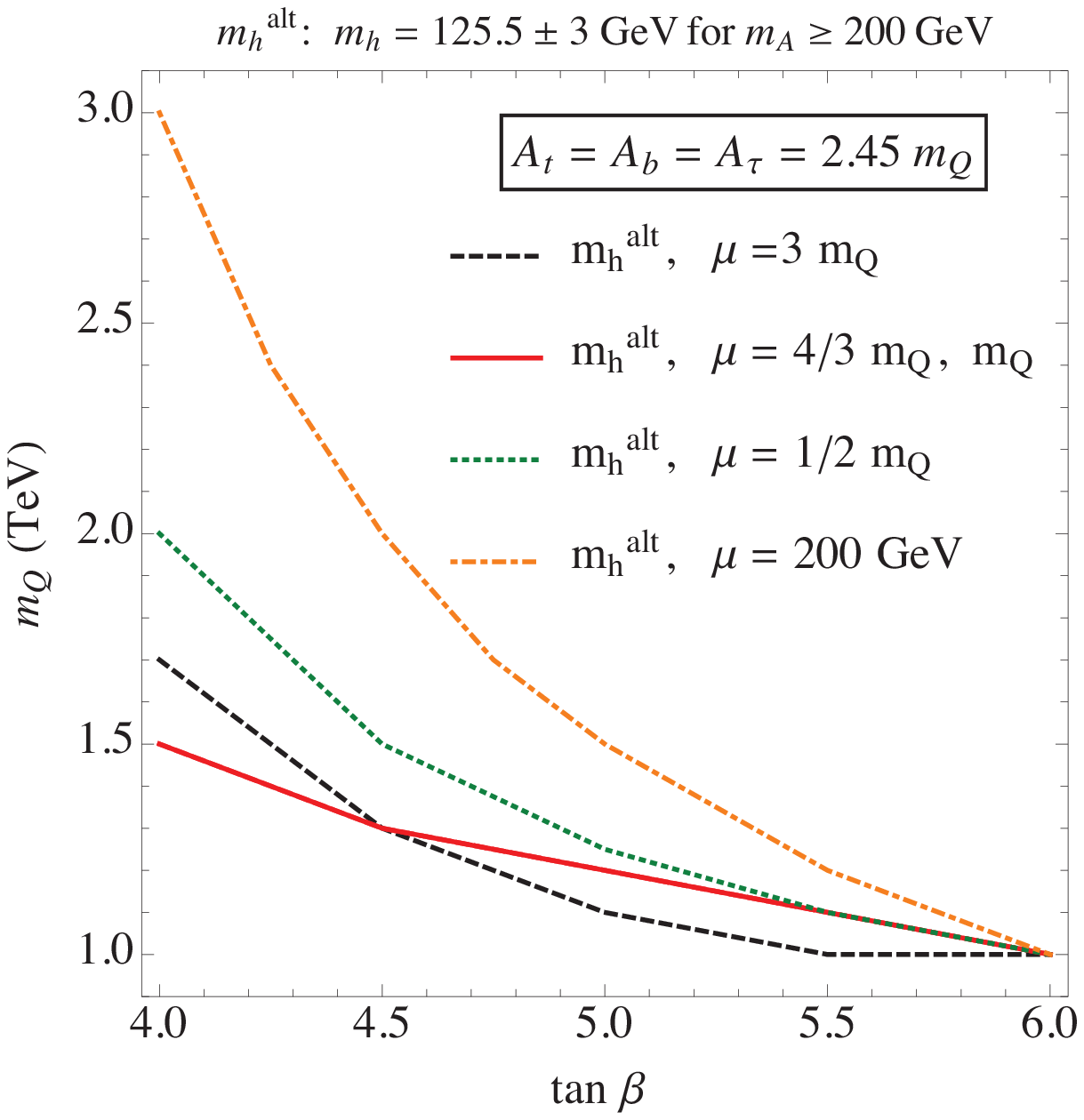}}~~
\subfloat[]{\includegraphics[width=3.2in, angle=0]{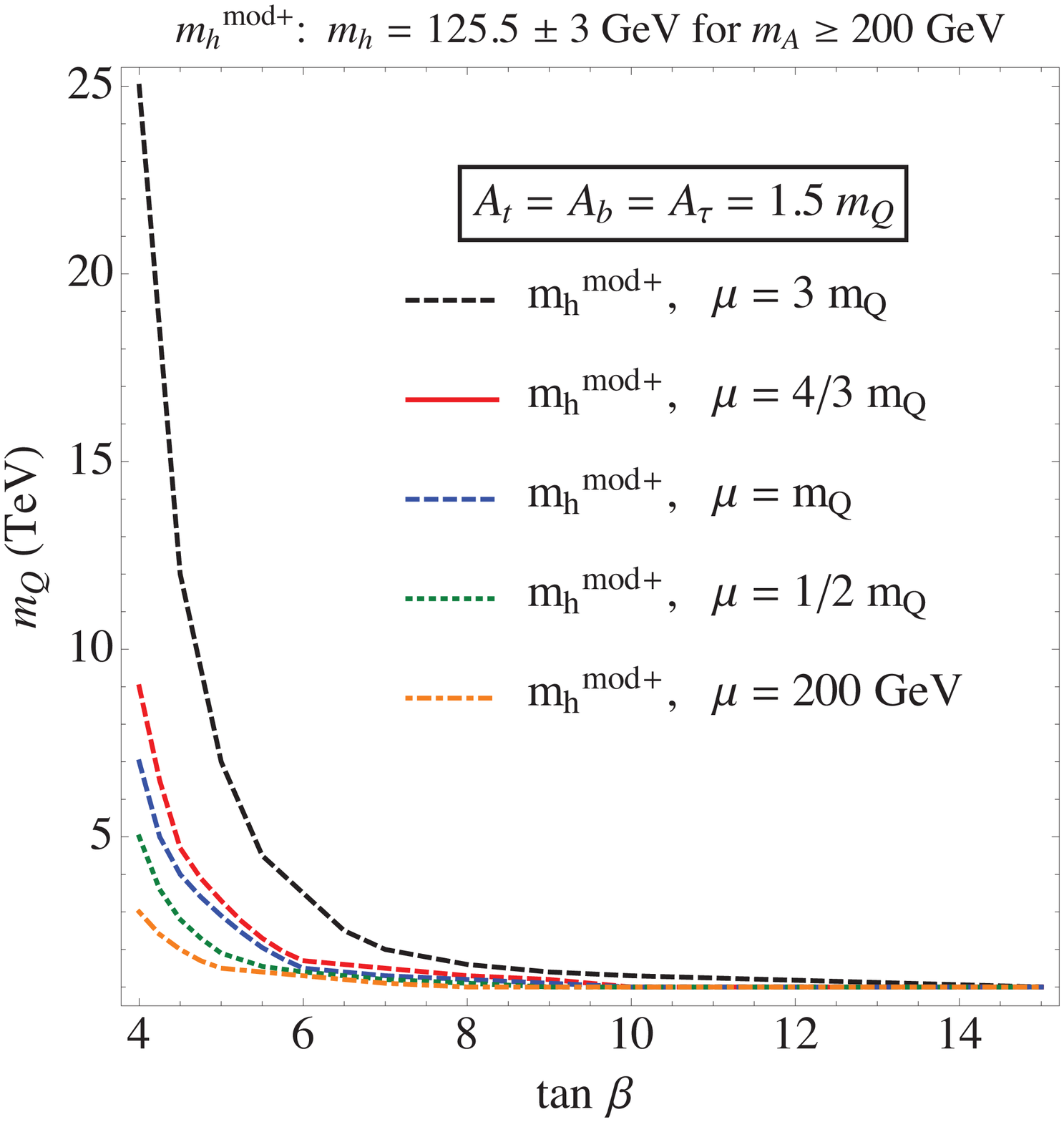}}
\caption{\label{fig:mQfromh}{\em Values of $m_Q$ necessary to accommodate the proper
value of the lightest CP-even Higgs mass, for different values of $\mu$ in the $m_h^{\rm alt}$ and $m_h^{\rm mod+}$ scenarios. }}
\end{figure}

The corresponding values of the stop soft breaking mass parameters, $m_Q$, are displayed in Fig.~\ref{fig:mQfromh}. Observe that for the $m_{h}^{\rm alt}$ scenario (apart for the case of $\mu=3 m_Q$), larger values of $m_Q$ are necessary for smaller values of $\mu$, while in the $m_{h}^{\rm mod+}$ scenario, larger values of $m_Q$ are obtained for larger values of $\mu$. The reason for this behavior is that generally in the $m_{h}^{\rm alt}$ scenario, larger values of $\mu$ approach the stop mixing for which the light CP-even Higgs mass is maximized, $X_t = A_t - \mu/t_\beta \simeq 2m_Q$, in the on-shell scheme. This implies the need for smaller logarithmic corrections, and therefore smaller values of $m_Q$.  The exception is the case of $\mu=3 m_Q$, where $\mu$ is so large that at small values of $t_\beta$, $X_t$ is already smaller than the maximal value for the Higgs mass. As $t_\beta$ increases, $X_t$ increases, approaching the maximal value from the other side. This explains the different dependence on $m_Q$ for this case. In the $m_{h}^{\rm mod+}$ scenario, larger values of $\mu$ imply values of $X_t$ further away from maximal mixing, which in turn require larger values of $m_Q$ to obtain the correct $m_h$.


\subsection{Decay Branching Ratios of Heavy Higgs Bosons}

\begin{figure}[p]
\subfloat[]{\includegraphics[width=3.3in, angle=0]{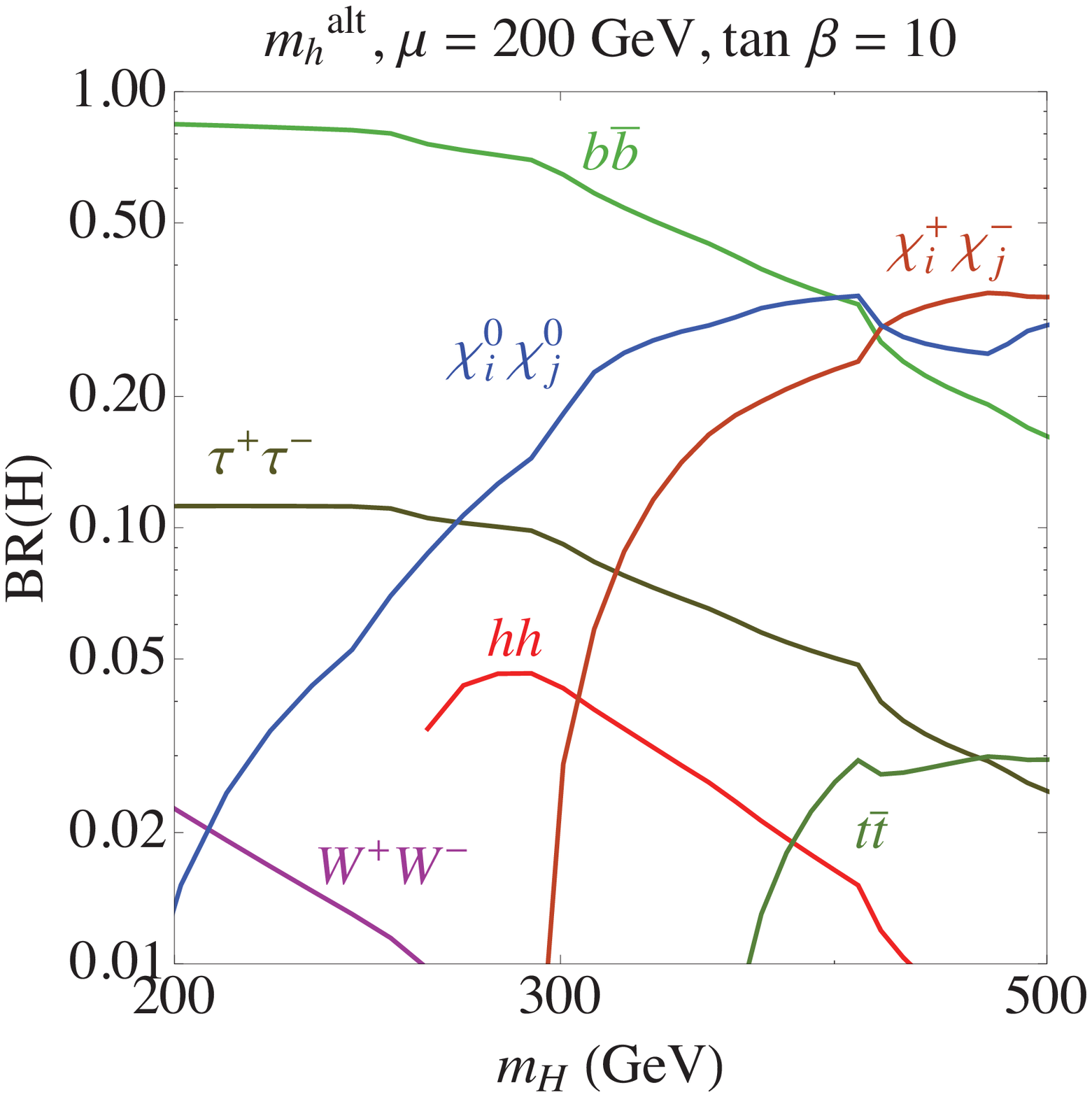}}~
\subfloat[]{\includegraphics[width=3.3in, angle=0]{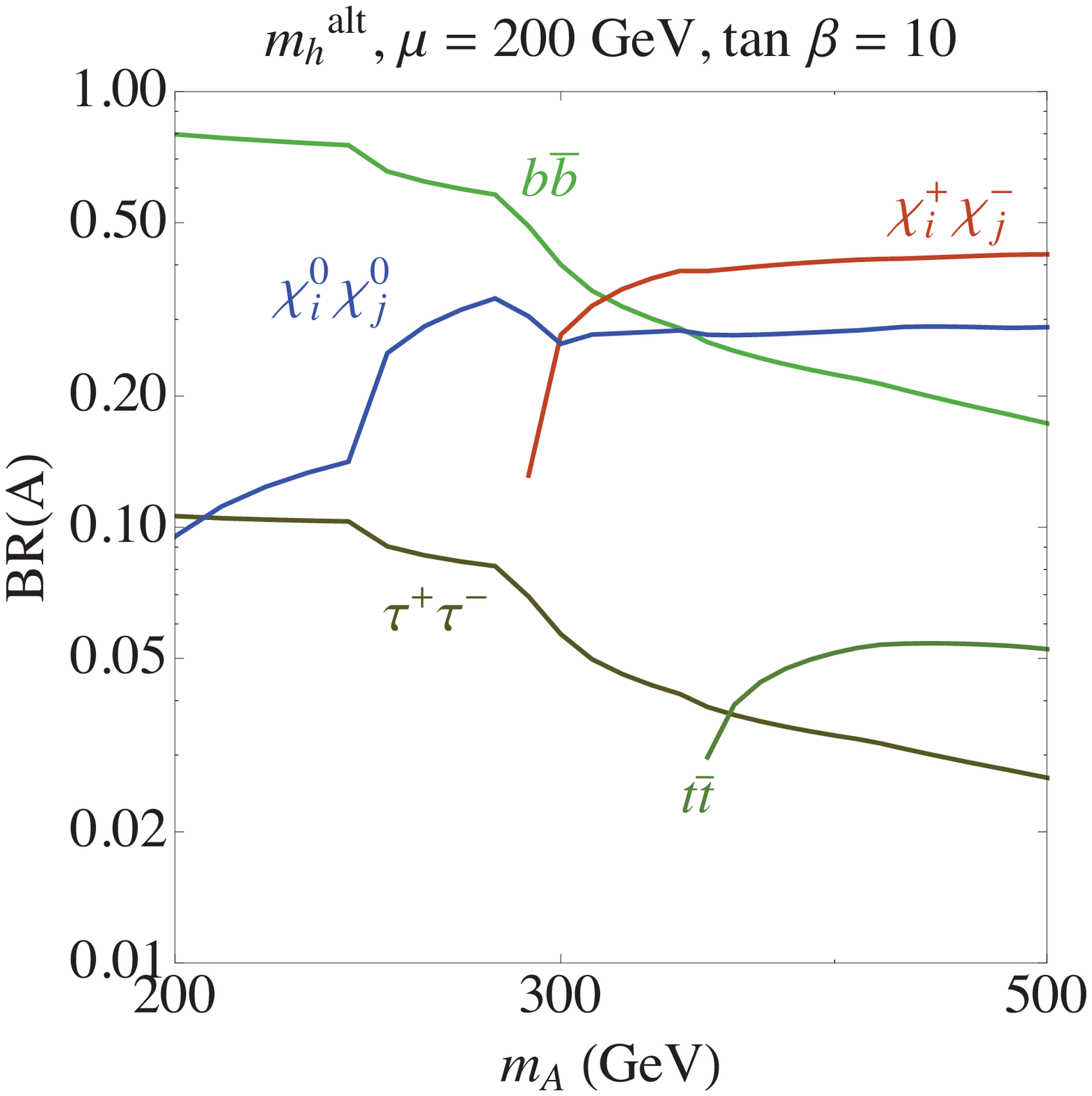}} \\
\subfloat[]{\includegraphics[width=3.3in, angle=0]{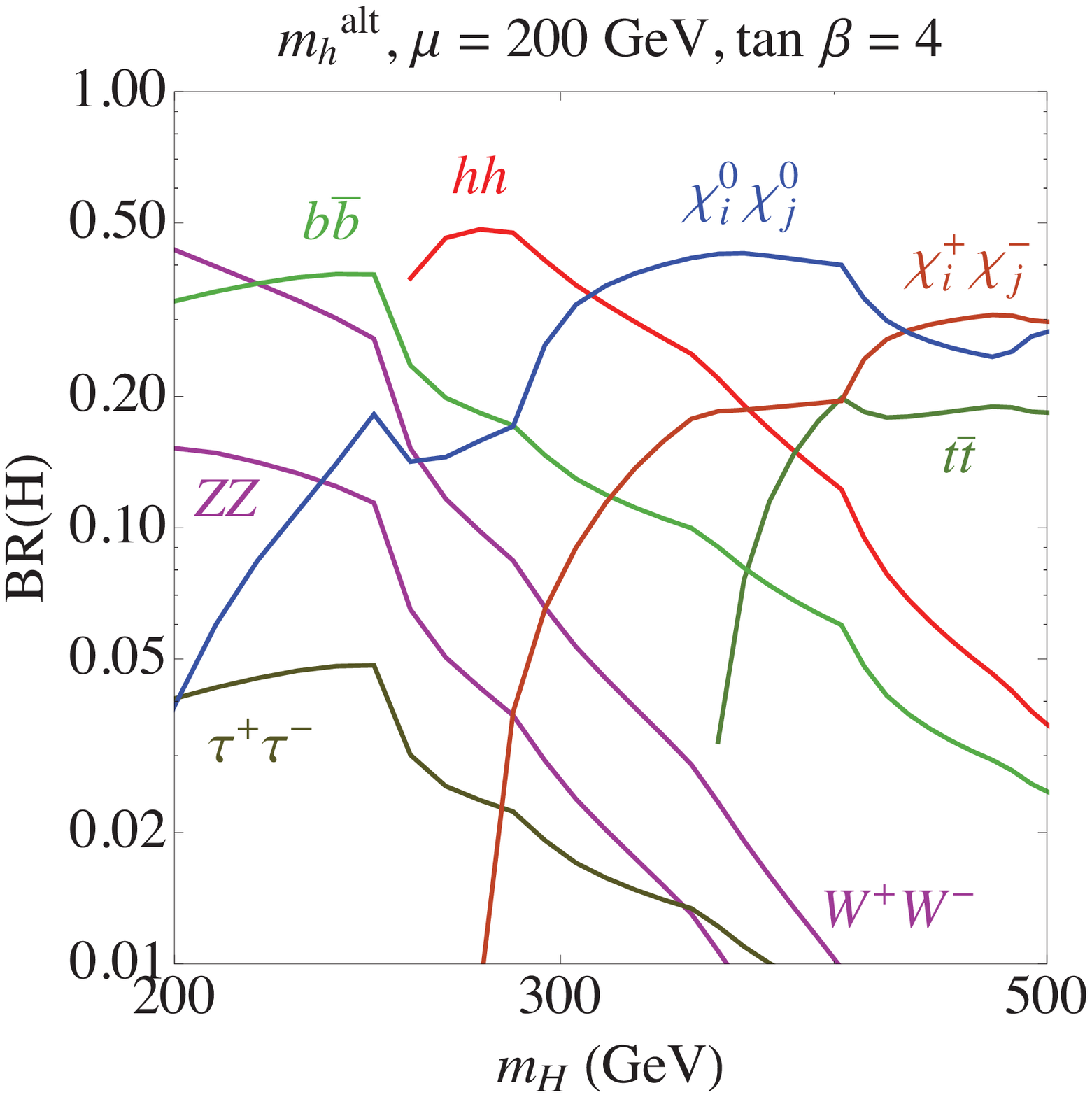}}~
\subfloat[]{\includegraphics[width=3.3in, angle=0]{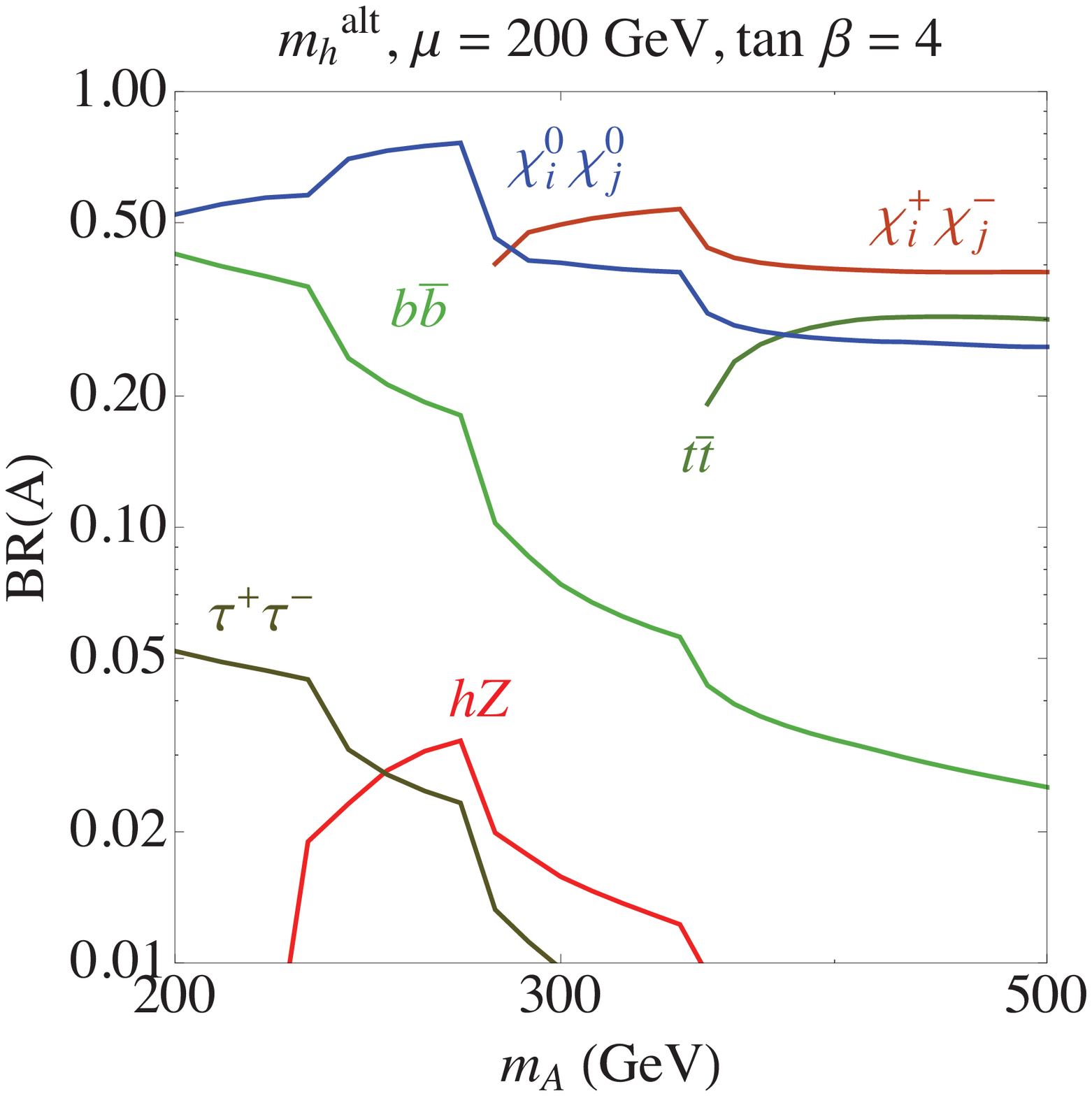}}
\caption{\label{fig:BRHAlowmu}{\em Branching Ratios of the heavy CP-even~(left panels) and CP-odd~(right panels) Higgs  bosons as a function of their respective masses in the m$_{\rm h}^{\rm alt}$ scenario, for $t_\beta = 10$~(top panels) and $t_\beta =4$~(bottom panels), for small values of the Higgsino mass parameter, $\mu = 200 \;{\rm GeV.}$}}
\end{figure}

\begin{figure}[p]
\subfloat[]{\includegraphics[width=3.3in, angle=0]{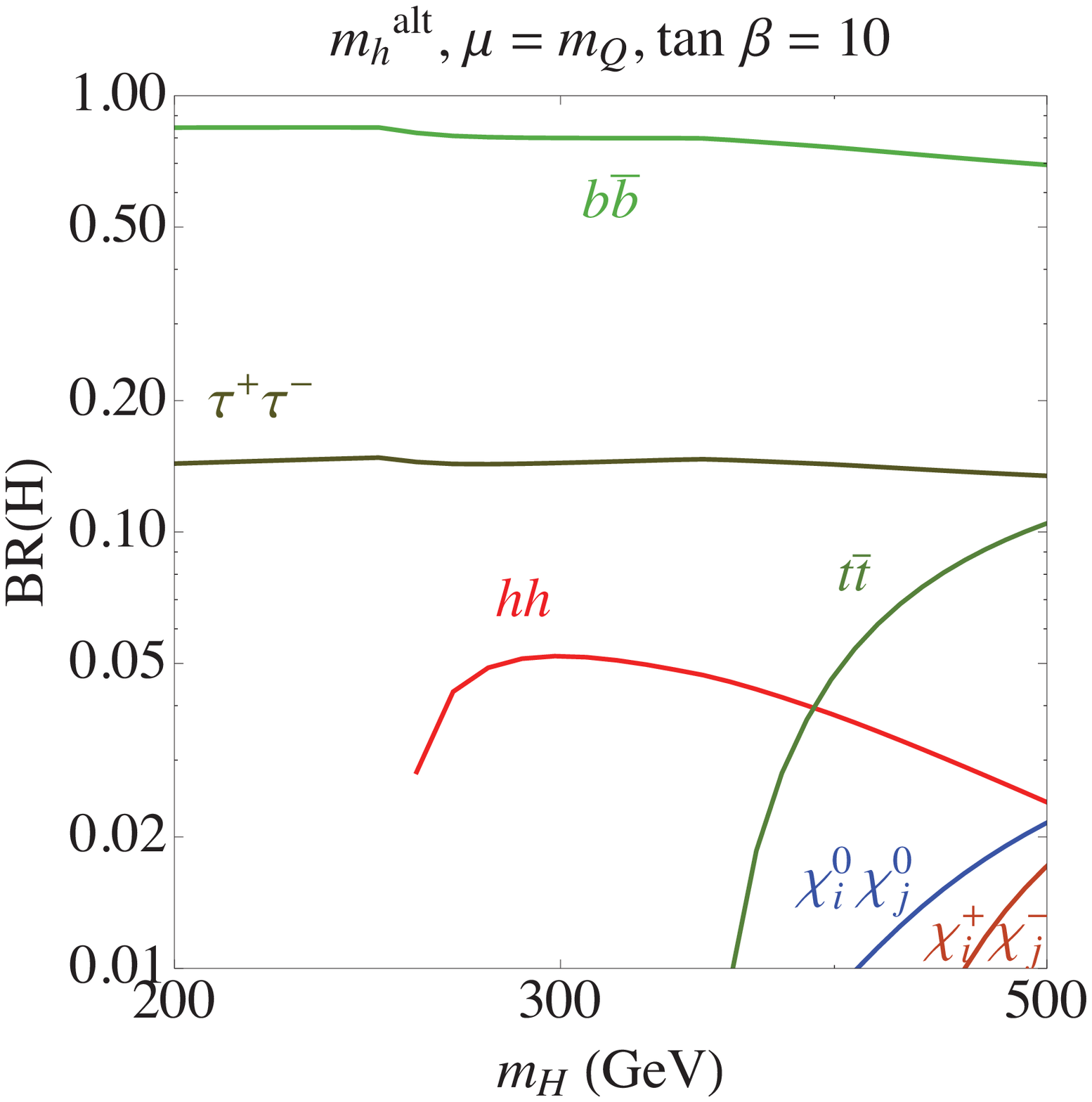}}~
\subfloat[]{\includegraphics[width=3.3in, angle=0]{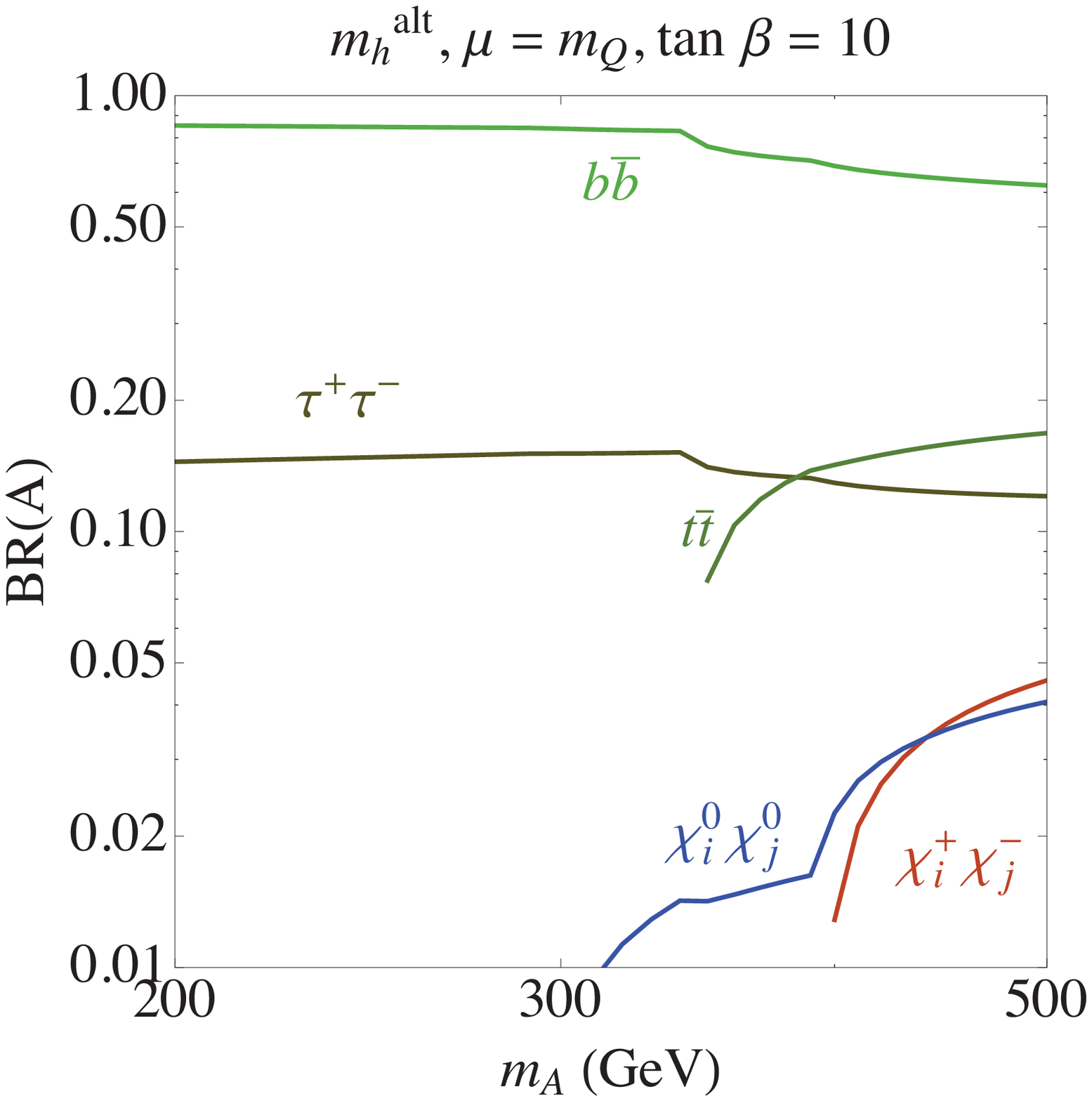}}  \\
\subfloat[]{\includegraphics[width=3.3in, angle=0]{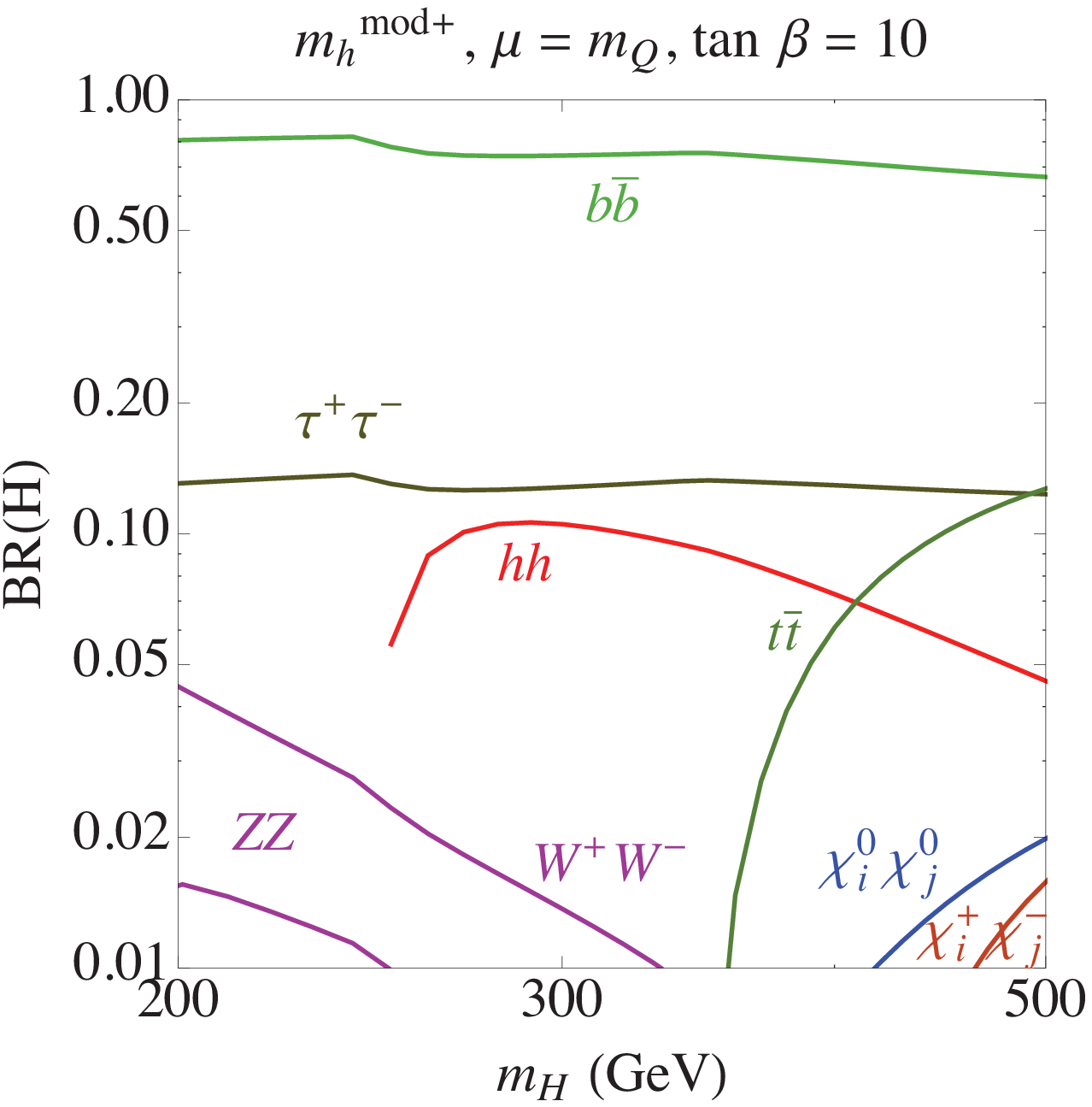}}~
\subfloat[]{\includegraphics[width=3.3in, angle=0]{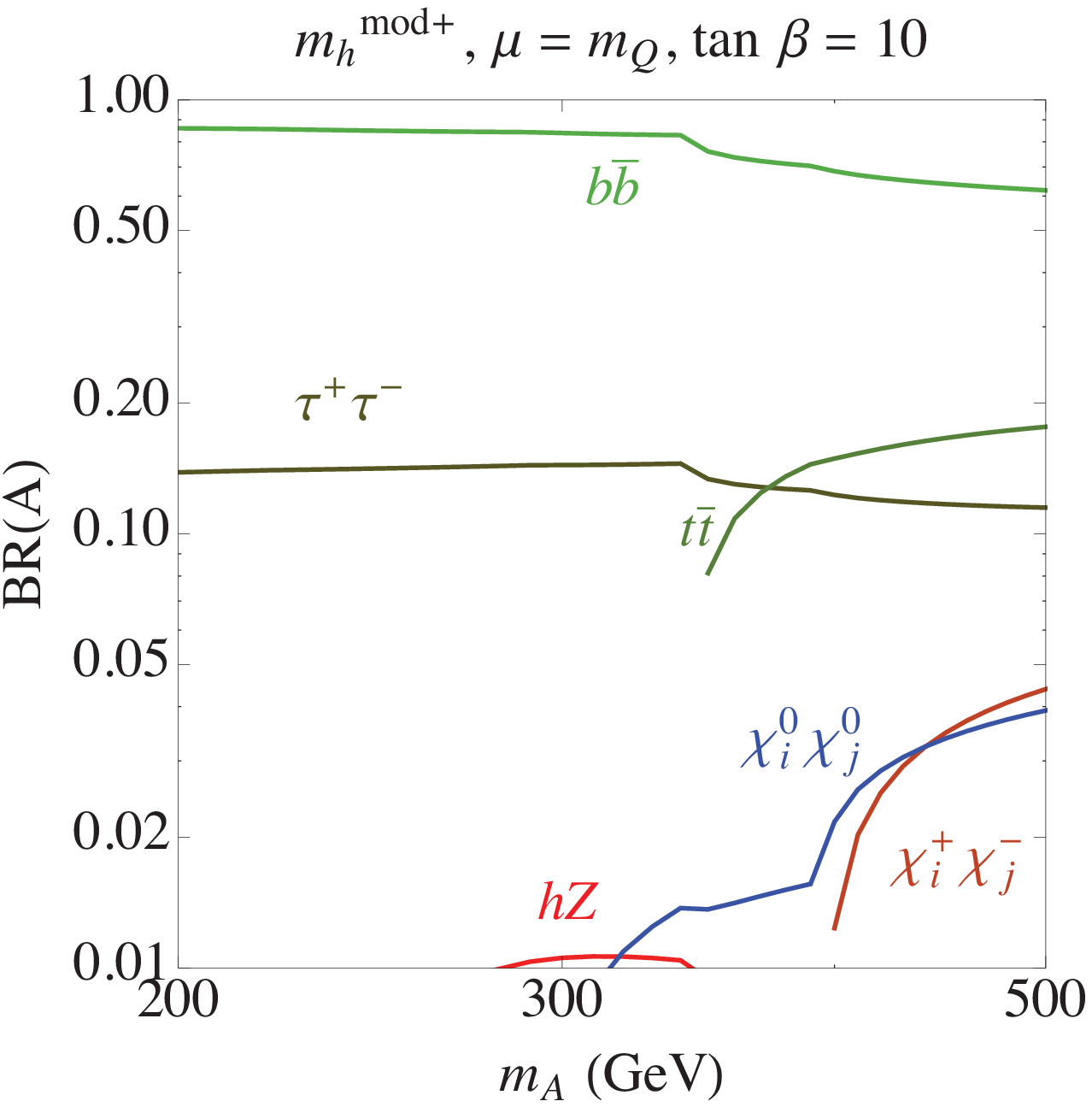}}
\caption{\label{fig:BRHAtanb10highmu}{\em  Branching Ratios of the heavy CP-even~(left panels) and CP-odd~(right panels) Higgs  bosons as a function of their respective masses for $t_\beta = 10$ in the m$_{\rm h}^{\rm alt}$ scenario~(top panels) and $m_h^{\rm mod+}$ scenario~(bottom panels), for large values of the Higgsino mass parameter, $\mu = m_Q$.}}
\end{figure}

\begin{figure}[p]
\subfloat[]{\includegraphics[width=3.3in, angle=0]{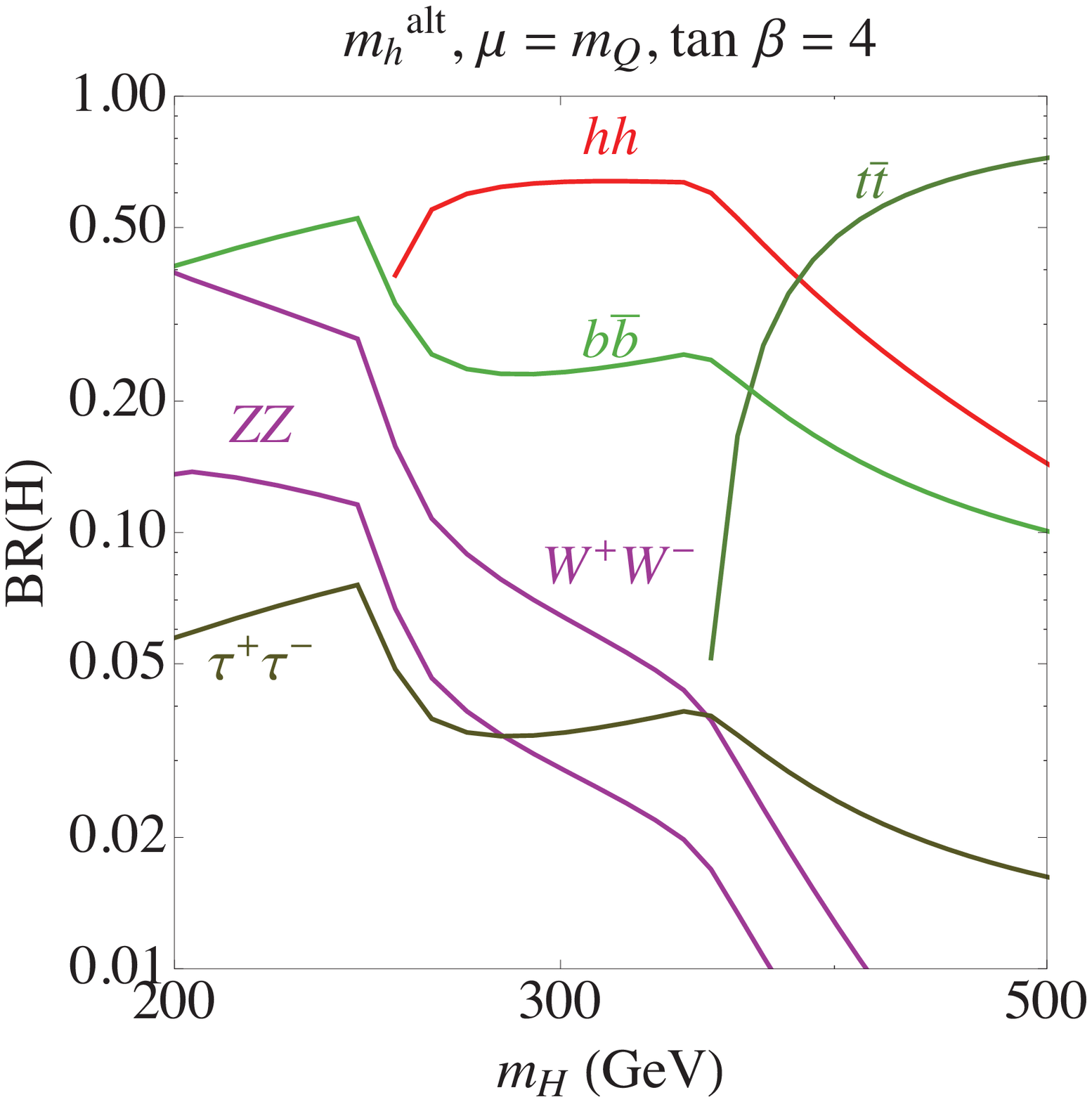}}~
\subfloat[]{\includegraphics[width=3.3in, angle=0]{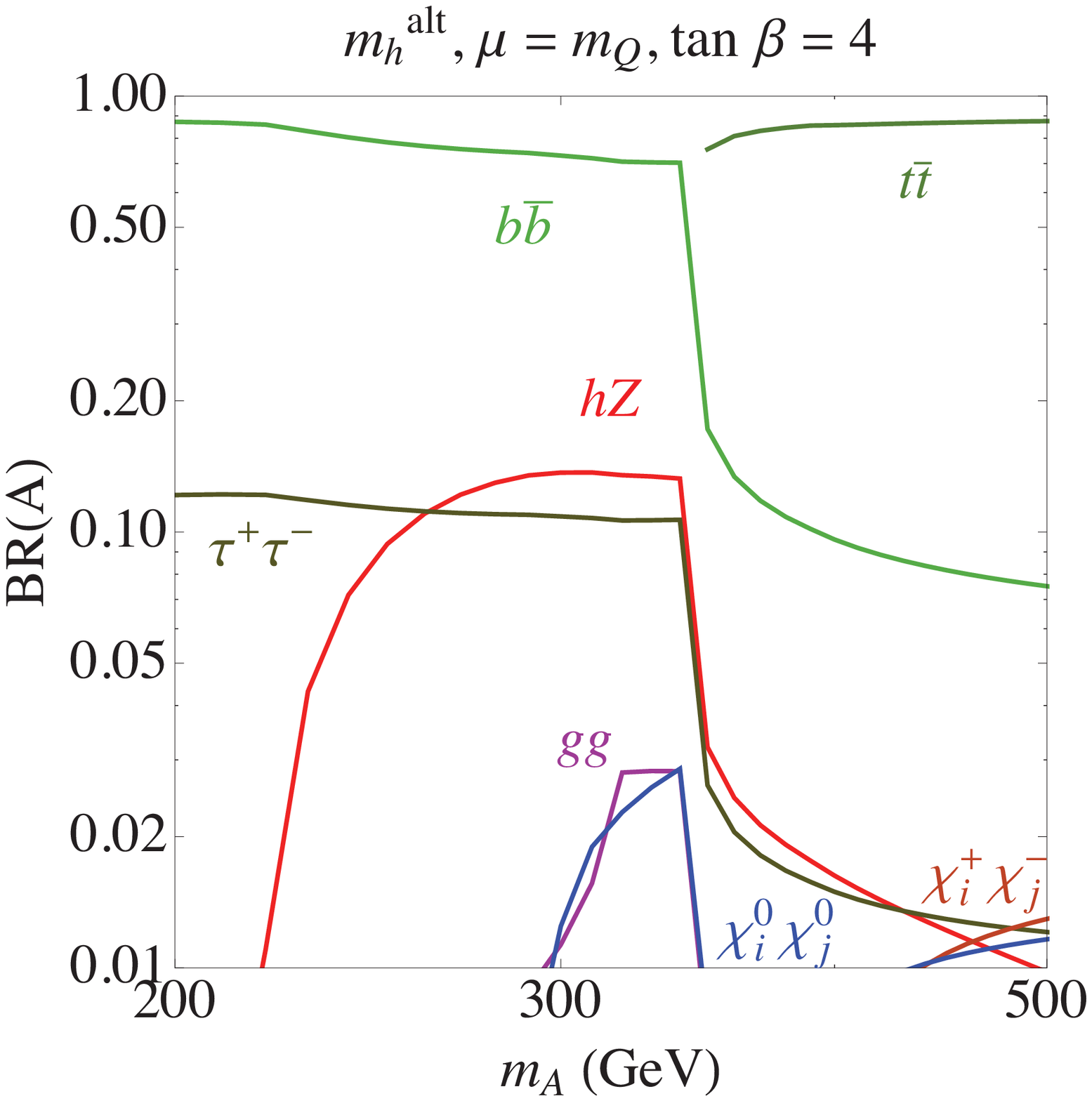}}  \\
\subfloat[]{\includegraphics[width=3.3in, angle=0]{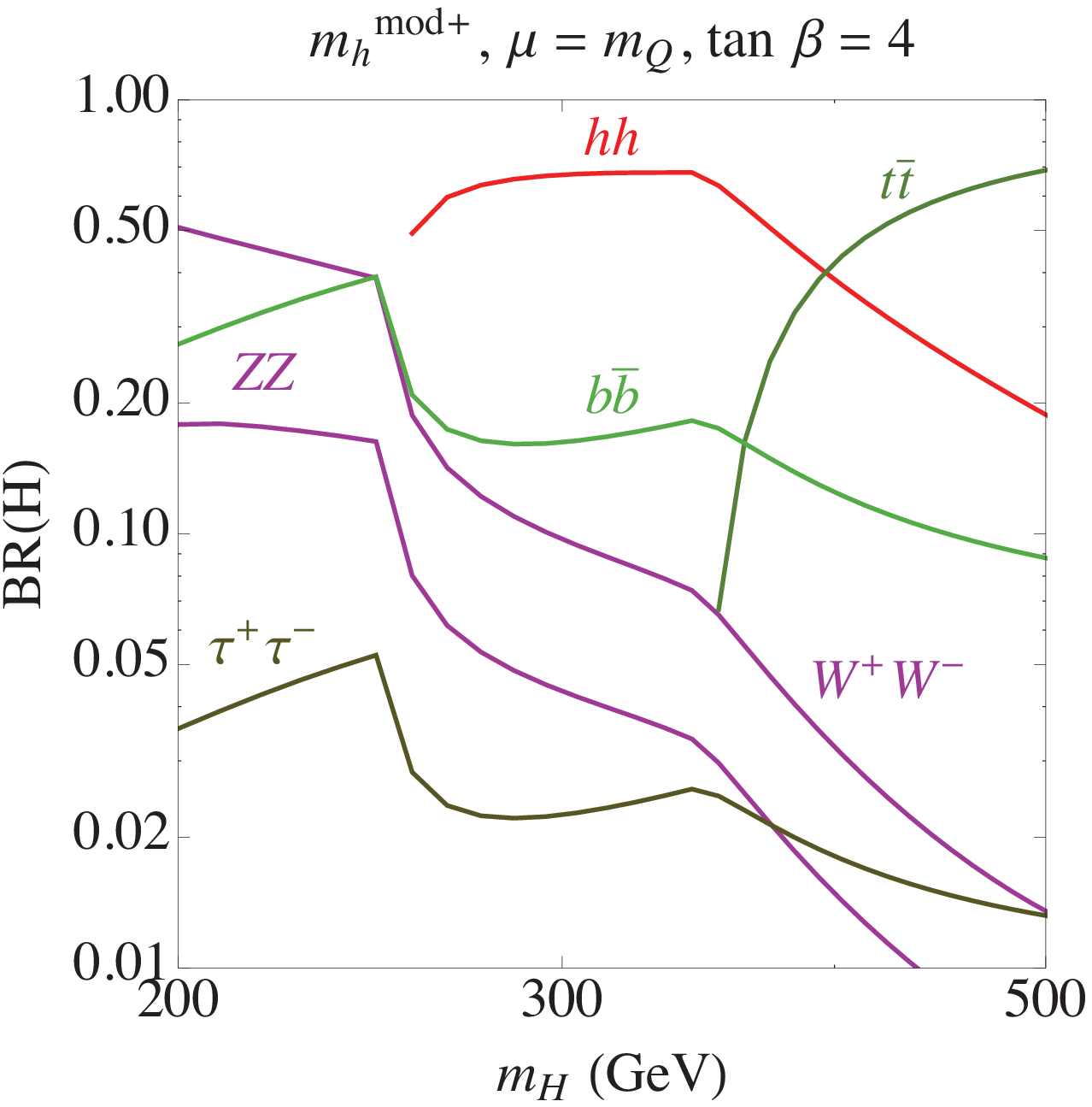}}~
\subfloat[]{\includegraphics[width=3.3in, angle=0]{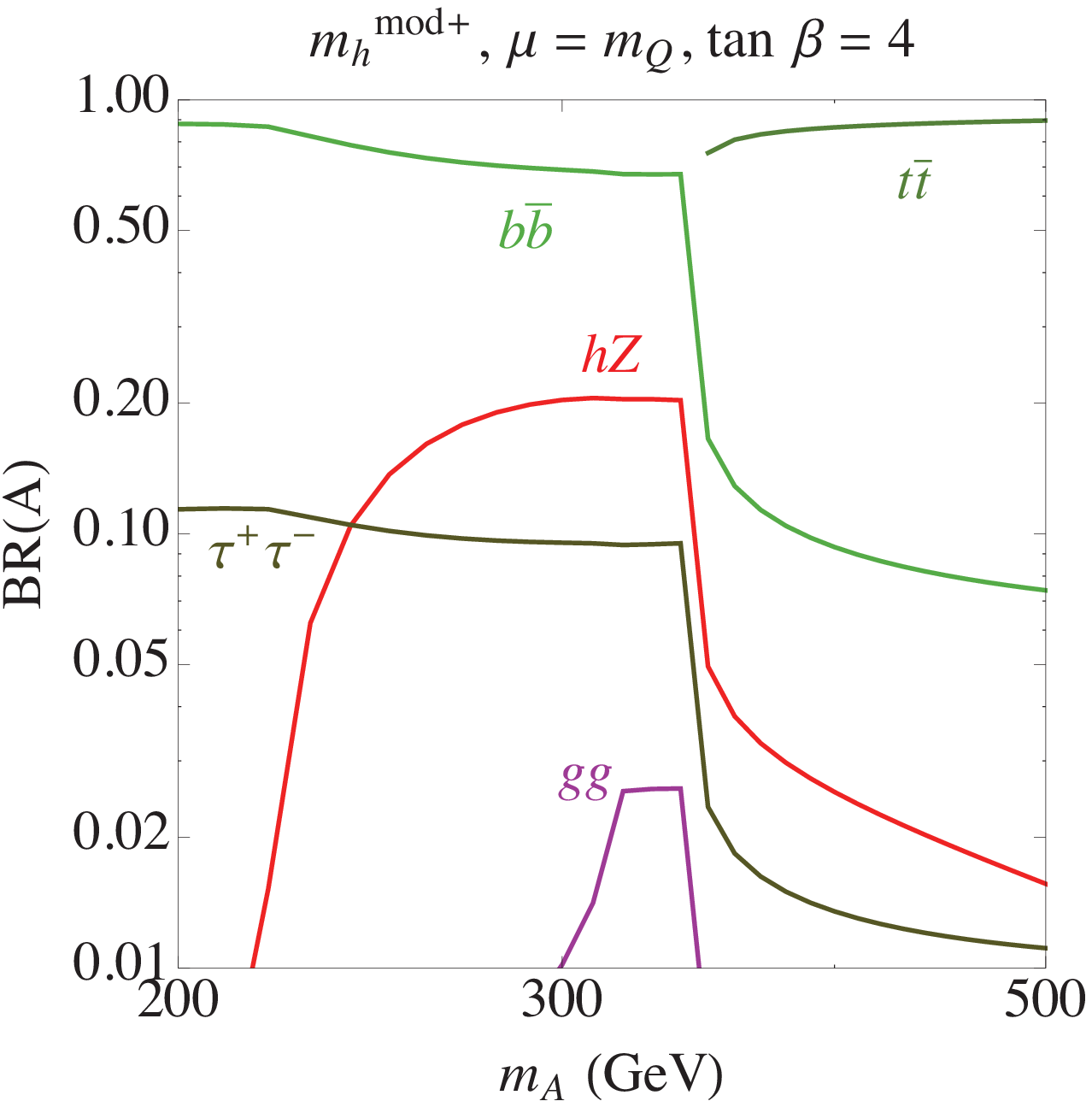}}
\caption{\label{fig:BRHAtanb4highmu}{\em  Branching Ratios of the heavy CP-even~(left panels) and CP-odd~(right panels) Higgs  bosons as a function of their respective masses for $t_\beta = 4$ in the m$_{\rm h}^{\rm alt}$ scenario~(top panels) and $m_h^{\rm mod+}$ scenario~(bottom panels), for large values of the Higgsino mass parameter, $\mu = m_Q$.}}
\end{figure}

In Fig.~\ref{fig:BRHAlowmu}  we show the variation in the decay branching ratios of the heavy
neutral Higgs bosons,  $H$ and $A$, in the $m_{h}^{\rm alt}$ scenario for small values of $\mu$,  and for moderate values  of $t_\beta = 10$ and small values of $t_\beta = 4$; the results in the $m_{h}^{\rm mod+}$ scenario for the same values of $\mu$ are very similar and will not be shown here.  At larger values of $\mu$, the distinction between the two scenarios becomes more prominent as shown in Figs.~\ref{fig:BRHAtanb10highmu} and \ref{fig:BRHAtanb4highmu}.

We first examine the case of small $\mu$. For $t_\beta = 10$, the decays into bottom-quarks represent the dominant decay mode of the heavy Higgs bosons at small values of $m_{A,H}$.  At the largest values of the non-standard Higgs boson masses shown in Fig.~\ref{fig:BRHAlowmu}, the decays of the heavy Higgs bosons into charginos and neutralinos become prominent, suppressing the branching ratio of the decays of the non-standard Higgs bosons into $b\bar{b}$ and $\tau^+\tau^-$.

For $t_\beta=4$, one interesting feature is that the decay of $H$ into pairs of lightest CP-even Higgs becomes significant at masses above the corresponding kinematic threshold, a property that  persists even when the value of $\mu$ is changed, as shown in Fig.~\ref{fig:BRHAtanb4highmu}. Another important feature is that the $H/A$ decay into pairs of neutralinos and charginos becomes prominent throughout the mass range we consider, thereby suppressing the decay branching ratios into the canonical search channels, $b\bar{b}$ and $\tau^+\tau^-$. In particular, the branching ratio of the heavy Higgs bosons into tau-lepton pairs, which is the main focus of present searches,  never exceeds 5\%  and is quite suppressed for $m_{A,H} \simgt 300 $~GeV.

Next we compare the decay branching ratios in the $m_h^{\rm mod+}$ and $m_h^{\rm alt}$ scenarios for large values of $\mu$. Fig.~\ref{fig:BRHAtanb10highmu} shows the comparison at $t_\beta=10$ while Fig.~\ref{fig:BRHAtanb4highmu} is for $t_\beta=4$. One important consequence of raising $\mu$ is  that the Higgsinos become heavy, resulting in small couplings of the light gaugino-like charginos and neutralinos to the neutral Higgs bosons. Therefore, the decays into electroweakinos are always suppressed, never exceeding a few percent. At $t_\beta = 10$ the decays into bottom-quark and tau-lepton pairs become prominent for all values of the heavy Higgs boson masses.

For $t_\beta =4$, the branching ratio of the decay of the heavy neutral Higgs bosons into bottom quarks and tau leptons is suppressed due to the decrease of the couplings of down-type fermions to these Higgs bosons. Hence, for $t_\beta=4$, the $H\to hh$ decay becomes the dominant mode for $m_H$ larger than the kinematic threshold of $2m_h$, until the top channel opens up and becomes the main decay mode. Even below the $2m_h$ threshold, the decay width of the heavy CP-even Higgs boson into weak gauge bosons is large enough to suppress the BR($H\to \tau^+\tau^-$) to values of order of 5\% in both scenarios. As for the CP-odd Higgs boson,  as can be seen in the right panels of Fig.~\ref{fig:BRHAtanb4highmu}, due to the absence of any relevant contribution to the total decay width beyond the bottom-quark and tau-lepton final states, the BR($A\to \tau^+ \tau^-$) remains of the order of 10\% up to the top quark pair decay threshold. It is worth noting that although the $hZ$ channel becomes significant when the kinematics allow,  for the same masses of the heavy Higgs bosons, BR$(A \to hZ)$ is always significantly lower than BR$(H \to hh)$.  These differences between the CP-even and CP-odd Higgs bosons have important phenomenological consequences that will be discussed below.

\subsection{Inclusive Production Rates of Heavy Higgs Bosons in the $\boldsymbol{\tau^+\tau^-}$ Channel}

At the LHC we only measure the total rate, i.e.~the production cross-section times the branching fraction into some specific final state. In particular, the strongest constraints in the MSSM on the $m_A -t_\beta$ plane are derived using searches in the $\tau^+\tau^-$ final states, which we focus on in this subsection. The main production modes for the heavy neutral Higgs bosons, $A$ and $H$, are the gluon fusion channel and, at moderate or large values of $t_\beta$,  associated production with bottom quarks. At large $t_\beta$, the main contribution to the gluon fusion cross section comes from bottom quark loops, since the heavy Higgs couplings to $b$-quarks are enhanced by $t_\beta$. Then the total production cross section is proportional to the square of the bottom Yukawa coupling. However, as $t_\beta$ decreases, the bottom coupling decreases while the top coupling to the non-standard Higgs bosons increases with $1/t_\beta$. Therefore, at values of $t_\beta \alt 6$, the dominant contribution to the gluon fusion production cross section is proportional to the square of the top coupling to the heavy neutral Higgs bosons and becomes significant.

\begin{figure}[t]
\subfloat[]{\includegraphics[width=3.2in, angle=0]{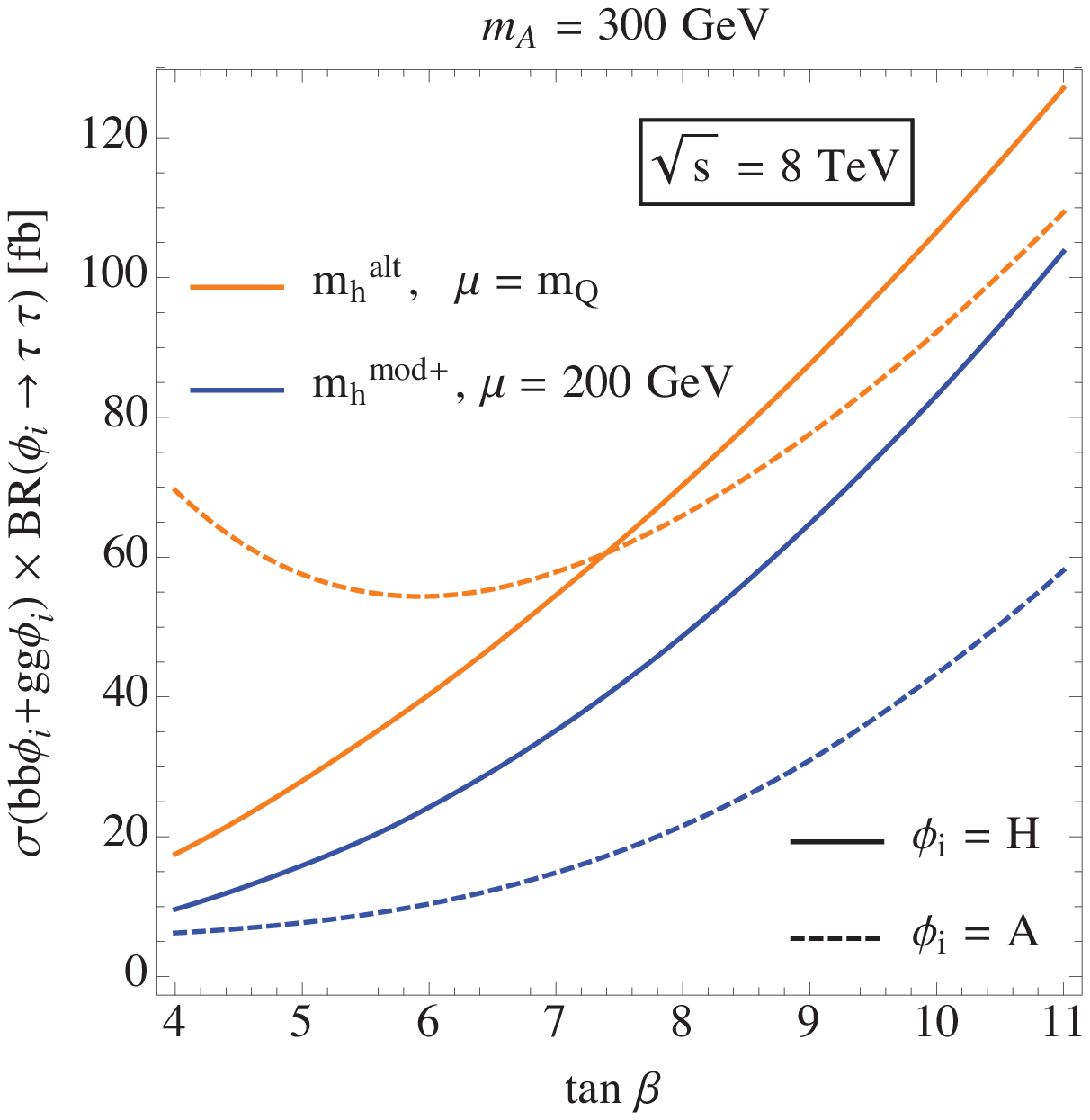}}~~~~
\subfloat[]{\includegraphics[width=3.2in, angle=0]{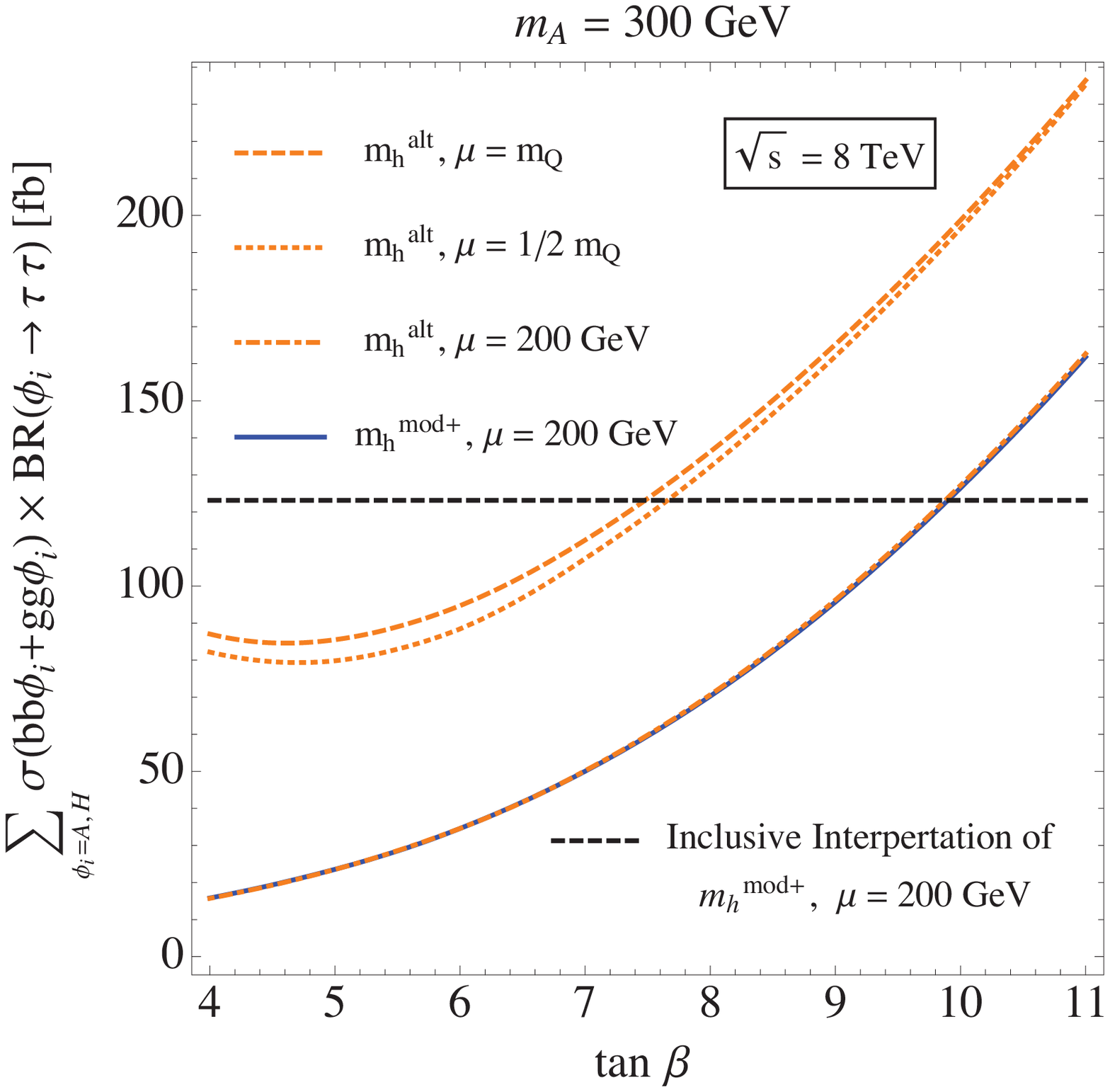}}
\caption{\label{fig:XSxBRmA300}{\em Inclusive production cross-section times branching ratio in the $\tau^+\tau^-$ mode for $m_A=300\;\rm{GeV}$. Black dashed line in right panel denotes extracted upper limit from CMS bounds presented in Ref.~\cite{Khachatryan:2014wca}.}}
\end{figure}

The left panel of Fig.~\ref{fig:XSxBRmA300} shows the dependence of the  inclusive production cross-section times the branching ratio of the decay of each neutral heavy Higgs boson into $\tau^+\tau^-$, for $m_A = 300$~GeV, in the $m_{h}^{\rm alt}$ and the $m_{h}^{\rm mod+}$ scenarios for different values of $\mu$. The solid lines display the behavior of the heavy CP-even Higgs boson and the dashed lines exhibit the corresponding CP-odd Higgs boson cross sections. The behavior of the Higgs-induced $\tau^+\tau^-$ production may be described using the properties of the production cross section and branching ratios discussed above. At large values of $\mu$, the CP-odd Higgs boson decay branching ratio into $\tau^+\tau^-$ remains large and approximately constant for all values of $t_\beta$, and hence the  total production rate into $\tau^+\tau^-$ closely follows the CP-odd Higgs production cross section. The increase of the production rate for the CP-odd Higgs boson into $\tau^+\tau^-$ at low values of $t_\beta$ and large $\mu$ is clearly seen in  Fig.~\ref{fig:XSxBRmA300}. Also visible is the fact that as $t_\beta$ decreases, the CP-even Higgs contribution to the $\tau^+\tau^-$ production rate is suppressed. This happens due to a decrease of the corresponding branching ratio, compensating for the increase in the gluon fusion production cross-section. The same happens for the CP-odd Higgs boson at low values of $\mu$.

The reach of the LHC in this channel  at low values of $t_\beta$ and $m_A = 300$~GeV becomes very different as one varies the $\mu$ parameter. For high values of $\mu$, the total production rate into $\tau^+\tau^-$ reaches a minimum at $t_\beta \simeq 6$ and then increases for lower values of $t_\beta$, as shown in the right panel of Fig.~\ref{fig:XSxBRmA300}. This is due to the CP-odd Higgs contribution as discussed above and shown in the left panel Fig.~\ref{fig:XSxBRmA300}. However, at low values of $\mu$, the inclusive production rate into $\tau^+\tau^-$ keeps decreasing for decreasing values of $t_\beta$, as also shown in the right panel of Fig.~\ref{fig:XSxBRmA300}.  The horizontal dashed line in the right panel of Fig.~\ref{fig:XSxBRmA300}  denotes an upper bound on the inclusive $\tau^+\tau^-$ production rate extracted from the CMS analysis in Ref.~\cite{Khachatryan:2014wca}~(the derivation and validity of this extracted limit is detailed in App. \ref{sect:bounds}). The value of $t_\beta$ where the horizontal dashed line meets the predicted cross-section, denotes the largest value of $t_\beta$ consistent with experimental observation. Values of $t_\beta$ above this should be considered ruled out because the inclusive production rate would be larger than the extracted upper bound. As more data is collected in Run II of the LHC, the bound on the $\tau^+\tau^-$ channel will become stronger and therefore the horizontal dashed line will be pushed towards smaller values if no scalar resonances are seen. If for a particular value of the mass of the heavy CP-even and CP-odd Higgs bosons the limit were pushed below the minimum of the inclusive $\tau^+\tau^-$ production rate in the large $\mu$ case, that particular value of the Higgs boson mass would  be excluded by the data for all values $t_\beta$. This is not possible for the low $\mu$ scenarios, for which no minimum of the production cross section is present.

At lower values of $m_A \simeq 200$~GeV the difference between low and high values of $\mu$ becomes less dramatic. Still, as can be seen from Figs.~\ref{fig:BRHAtanb10highmu} and \ref{fig:BRHAtanb4highmu}, at $t_\beta =  4$, BR($A\to \tau^+\tau^-$) remains of order $10\%$ for large values of $\mu$ and becomes about half of that value for low values of $\mu$. In contrast, BR($H\to \tau^+\tau^-$) is always somewhat suppressed due to the presence of the decay of the heavy CP-even Higgs into $VV$, suffering an additional suppression at low values of $\mu$. In this particular example at $t_\beta = 4$, BR($H\to \tau^+\tau^-$)  is of order $6\%$ for high values of $\mu$ and is reduced to about $3\%$ for low values of $\mu$. Hence, in this case the largest $\tau^+\tau^-$ production contribution comes from the CP-odd Higgs boson.

\subsection{Rescaling Current LHC limits}

\begin{figure}[tb]
\subfloat[]{\includegraphics[width=5in, angle=0]{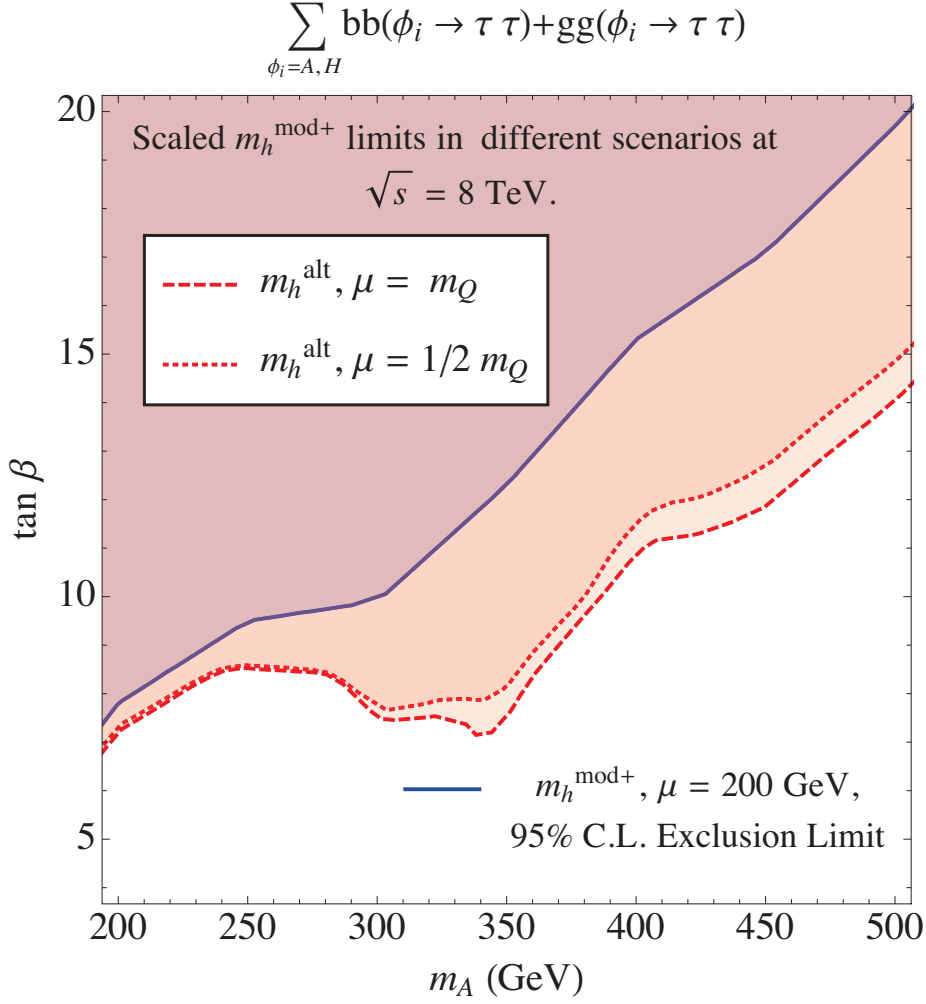}}
\caption{\label{fig:XSxBRscale}{\em Direct search bounds from the inclusive  $\tau^+\tau^-$ mode in our benchmarks at LHC8. The solid line displays the current CMS bounds in the m$_{\rm h}^{\rm mod+}$ scenario with $\mu=200$~{\rm GeV~\cite{Khachatryan:2014wca}}.}}
\end{figure}

We use the procedure discussed in App.~\ref{sect:bounds} to convert the $m_A$--$t_\beta$ limits presented by the experimental collaborations for a specific scenario, into limits on the inclusive production rate into $\tau^+\tau^-$ for a given value of $m_A$. We then demand that any other scenario we are considering leads to an inclusive production rate which is smaller than this extracted limit. In this way, we are able to obtain a simple rescaling algorithm for the values of $t_\beta$ excluded in any given scenario. The outcome of such a  procedure is presented in Fig.~\ref{fig:XSxBRscale}, which shows the exclusion limits on the $m_A$--$t_\beta$ plane in our $m_{h}^{\rm alt}$ scenario for two different choices of the $\mu$ parameter.  As stressed in the last section, an important distinction in going from small to large values of $\mu$ is that the Higgsinos become heavy and therefore the decays of the heavy Higgs bosons into neutralino and/or chargino pairs are suppressed, resulting in a larger branching fraction into $\tau^+\tau^-$ channels.
It is clear that, due to the increase in the $\tau^+\tau^-$ production rate for larger values of $\mu$ (see Fig.~\ref{fig:XSxBRmA300}), the exclusion limit may be extended to smaller values of $t_\beta$.

As previously noted, the existence of a minimum in the inclusive production rate for the $\tau^+\tau^-$ channel as a function of $t_\beta$ for large values of $\mu$ (cf.~Fig.~\ref{fig:XSxBRmA300}), means that if this minimum falls below the experimental upper bound in the future, one would exclude all $t_\beta$ for a particular value of $m_A$ in the scenario under consideration. Indeed, in Ref.~\cite{Arbey:2013jla} it was shown that for heavy supersymmetric particles, the LHC has the capability of probing the wedge region by means of the $H,A \to \tau^+\tau^-$ channel in the 14~TeV run.  However, since this minimum does not exist for the low $\mu$ scenarios, even at 14~TeV, it is unlikely that the LHC would be able to completely probe the low $m_A$--$t_\beta$ region for these cases.

\begin{figure}[tb]
\subfloat[]{\includegraphics[width=5in, angle=0]{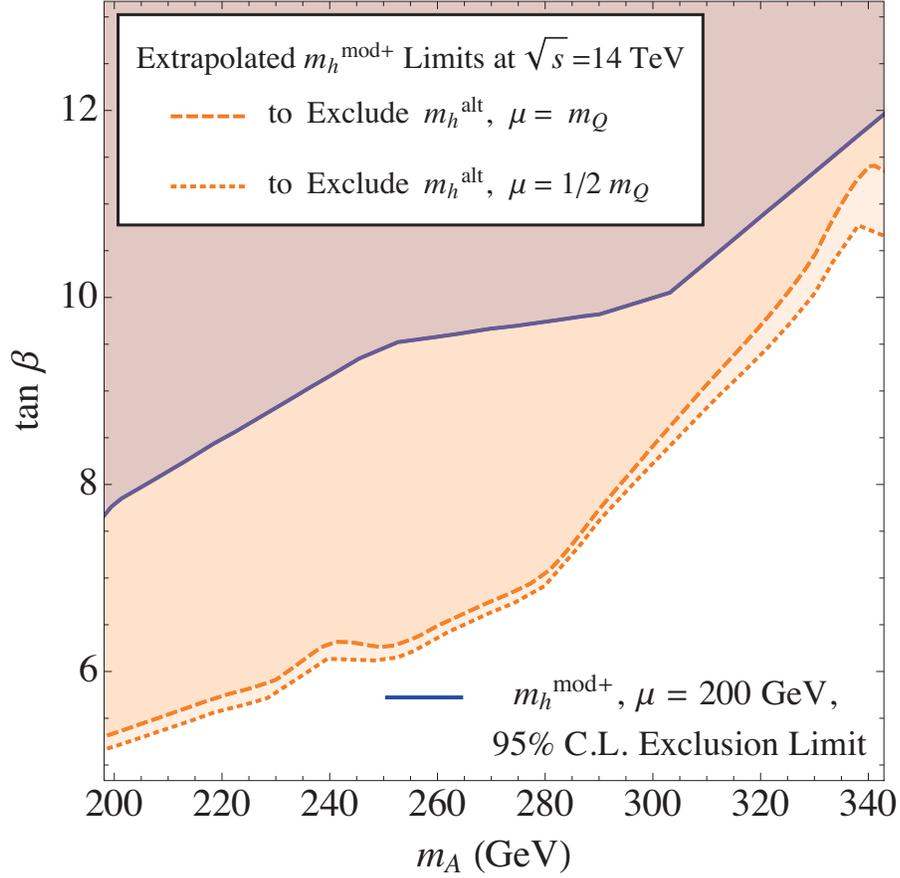}}
\caption{\label{fig:XSxBRneed}{\em The dashed and dotted line exhibit the projected bounds at $\sqrt{s}=14$~{\rm TeV} in the m$_{\rm h}^{mod+}$ scenario with $\mu = 200$~{\rm GeV}, such that all values of $t_\beta$ are excluded in  the m$_{\rm h}^{\rm alt}$ scenario for large values of $\mu$.  The solid line displays the current CMS bounds in the m$_{\rm h}^{\rm mod+}$ scenario with $\mu=200$~{\rm GeV~\cite{Khachatryan:2014wca}}.}}
\end{figure}

In Fig.~\ref{fig:XSxBRneed}  we show the projected limits in the $m_{ h}^{\rm mod+}$ scenario, with $\mu = 200$~GeV, that are required to exclude all values of $t_\beta$ in scenarios with large $\mu$ for $m_A <350$ GeV. More explicitly, if in the future the exclusion limit in the  $m_{ h}^{\rm mod+}$ scenario, with $\mu = 200$~GeV, reaches the dashed~[dotted] lines, the $m_h^{\rm alt}$ benchmark, with $\mu=m_Q\;[m_Q/2]$, would be completely ruled out, respectively. The situation for all our benchmarks with other choices of $\mu$ is similar, as long as $\mu \sim {\cal O}(m_Q)$ or larger. For comparison, the solid line in  Fig.~\ref{fig:XSxBRneed} represents the current bound from the LHC8 in Ref.~\cite{Khachatryan:2014wca} .

Note that in this article we have assumed that all squark masses are of the order of the stop masses, and hence the  next-to-lightest neutralinos and the lightest charginos would mostly decay into
the lightest neutralino and $Z$, $h$ and $W^\pm$, respectively.  Under these conditions, the values of $\mu$, $M_2$ and $M_1$ associated with the low $\mu$ scenario here are at the edge of the
current region of parameters probed by the ATLAS and CMS experiments~\cite{Aad:2014nua,Khachatryan:2014mma}.  Since the heavier Higgs bosons decay prominently into these particles, it would be interesting to perform a search for these
Higgs bosons decaying into charginos and neutralinos.   These will lead to final states already present in the decays of the heavier Higgs bosons into SM particles, namely $hh$, $VV$ and $Zh$,  that
are being studied at present (see, e.g., Refs.~\cite{Aad:2014yja, CMS:2014ipa}), but will be characterized by large amounts of missing energy.

\section{Precision $\boldsymbol{h}$ Measurements versus $\boldsymbol{H}$ and $\boldsymbol{A}$ Direct Searches.}

After analyzing the direct search constraints in the two classes of benchmarks with a varying $\mu$ parameter, we now study the interplay between direct searches and measurements of properties of the lightest CP-even Higgs boson at 125 GeV. The value of $\mu/m_Q$ plays an important role in determining the value of $t_\beta$ at which alignment  occurs, as can be seen in Eq.~(\ref{tbsol}). We shall show that the low $t_\beta$ and low $m_A$ region, which is difficult to probe in direct searches at low values of $\mu$,  results in deviations in the properties of the 125 GeV Higgs boson that are quite significant. Therefore, direct searches and precision Higgs measurements are complementary to each other.

In studying properties of the lightest CP-even Higgs boson, we will focus on its couplings to massive gauge bosons $h\to VV$, which are measured quite well experimentally. Another possibility is to use loop-induced couplings such as the diphoton coupling. Indeed,  the different values of $A_t$ and $\mu$ chosen in the $m_{h}^{\rm mod+}$ and $m_{h}^{\rm alt}$ scenarios lead to deviations in the loop-induced couplings. However, as is demonstrated in App.~\ref{appen}, the constraining power between these two couplings does not differ significantly.

It is worth emphasizing again that in order to study the complementarity between precision measurements and direct searches, it is important to obtain the correct mass for the lightest CP-even Higgs boson, which has a major impact on the properties of the 125 GeV Higgs boson and on the decays of the heavy Higgs bosons. As we showed in Section~\ref{sect:higgsmass}, in the region of interests where both $t_\beta$ and $m_A$ are small, the value of $m_Q$ should be raised to values larger than 1~TeV in order to obtain the proper lightest CP-even Higgs mass values.

Under the assumption of $|c_{\beta-\alpha}|\ll 1$, it follows from the results of Section~\ref{sec:MSSMHiggs} that
\be
g_{hVV}  \simeq  g_{hVV}^{\rm SM}\,,\qquad\quad
g_{htt}  \simeq  g_{ htt}^{\rm SM}\,,
\ee
whereas
\be \label{ghdd}
g_{hbb} \simeq g_{hbb}^{\rm SM} \left( 1 - c_{\beta-\alpha } t_\beta \right)\,,
\ee
where for simplicity we have neglected the $\Delta_b$ and $\delta h_b$ effects in \eq{hlbb}. This implies that, apart from small corrections coming from the squark loops contributing to the gluon-gluon fusion production, the lightest CP-even Higgs production cross section is SM-like.  Moreover, the decay branching ratios of the lightest CP-even Higgs boson are mostly affected by the modification of the bottom and $\tau$ couplings.  Inspection of \eq{tanbcbma} reveals that the down-type quark (and lepton) Yukawa couplings can significantly deviate from their corresponding SM values at low $m_A$ and moderate values of $t_\beta$. Moreover, for small values of $\mu$ these modifications are only weakly dependent on $t_\beta$, while for large values of $\mu$, a dependence on $t_\beta$ appears that may lead to alignment for the specific value of $t_\beta$ at which $c_{\beta-\alpha} = 0$.

\begin{figure}[p]
\subfloat[]{\includegraphics[width=3.3in, angle=0]{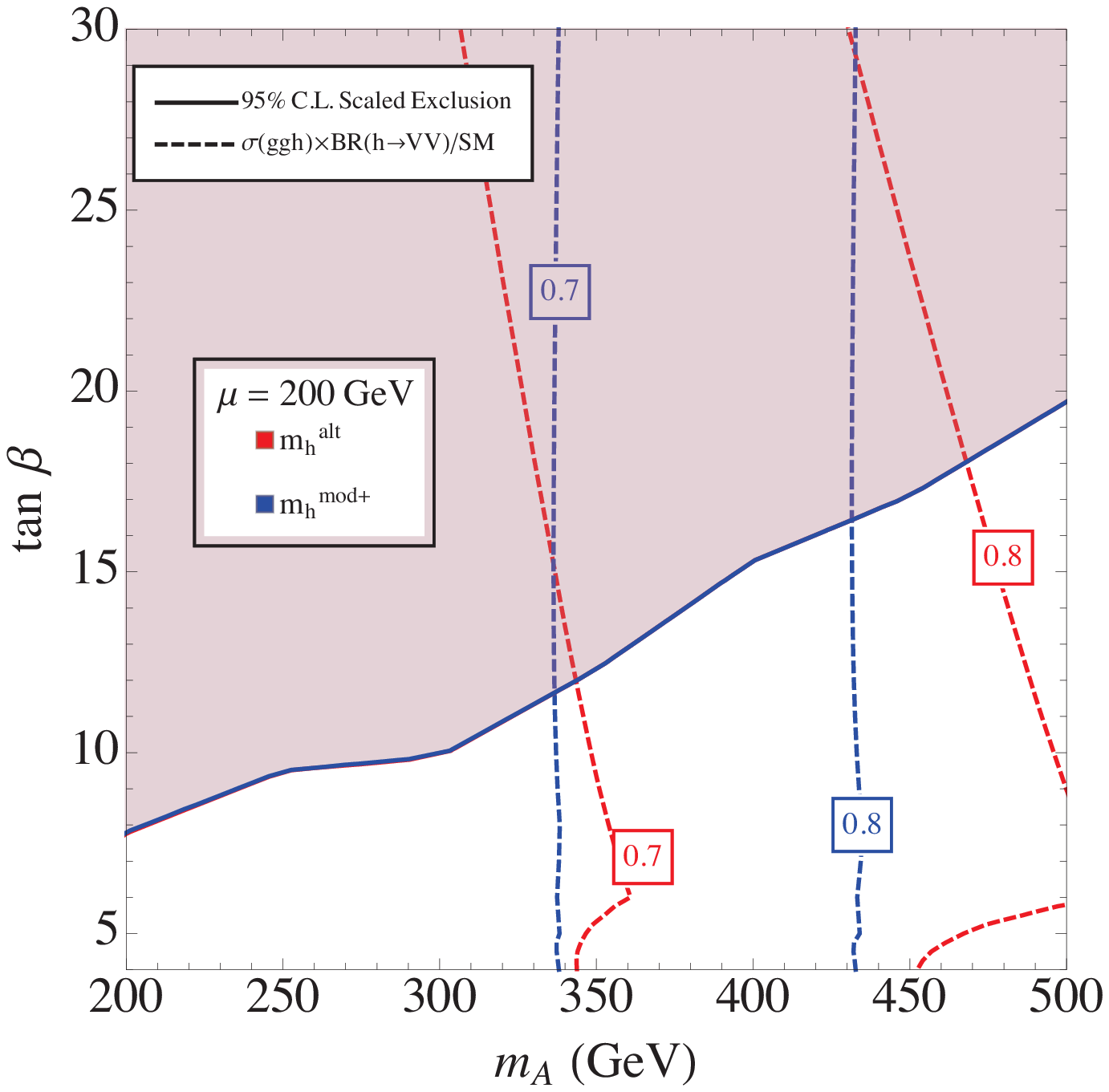}}~
\subfloat[]{\includegraphics[width=3.3in, angle=0]{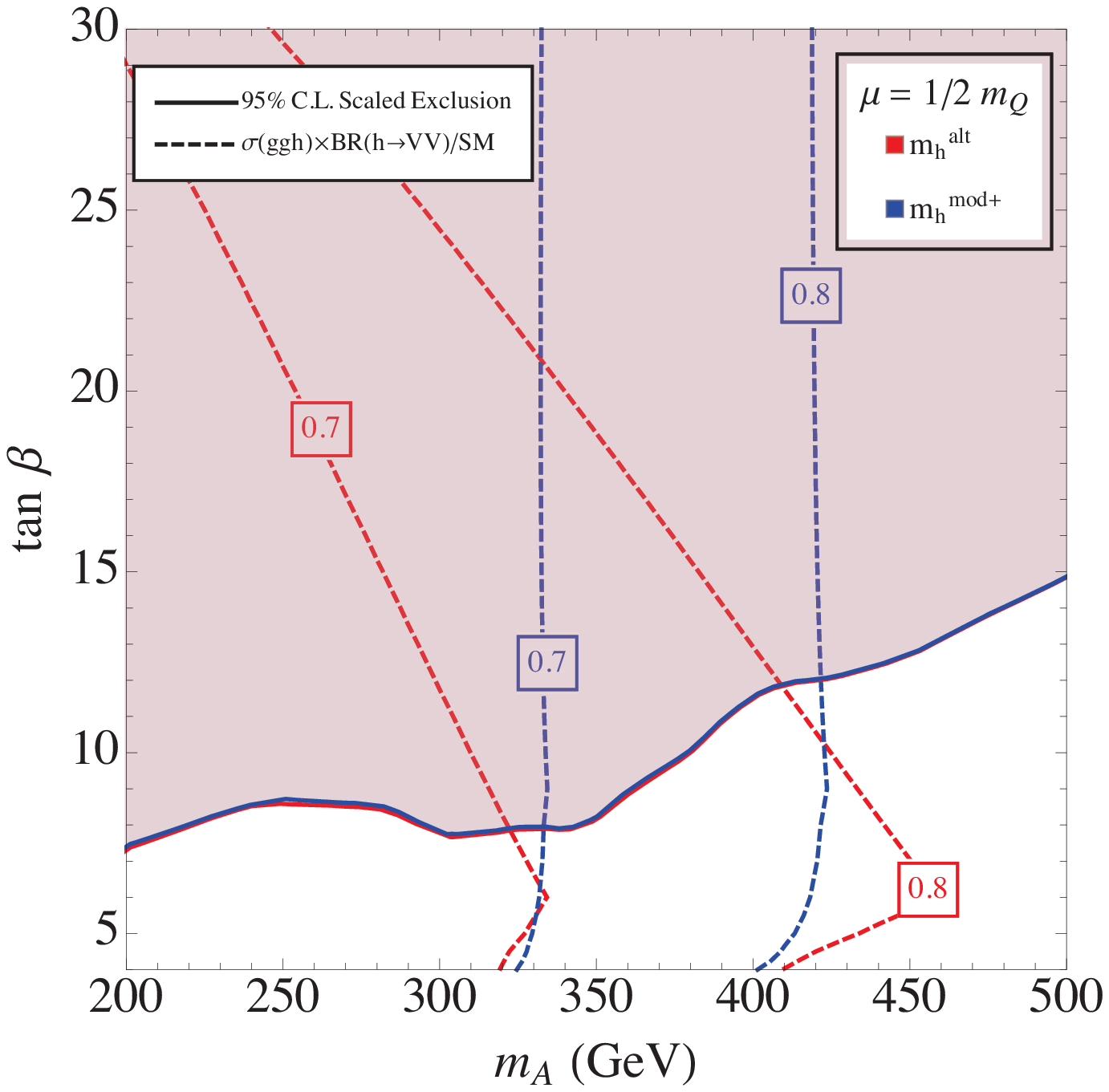}} \\
\subfloat[]{\includegraphics[width=3.3in, angle=0]{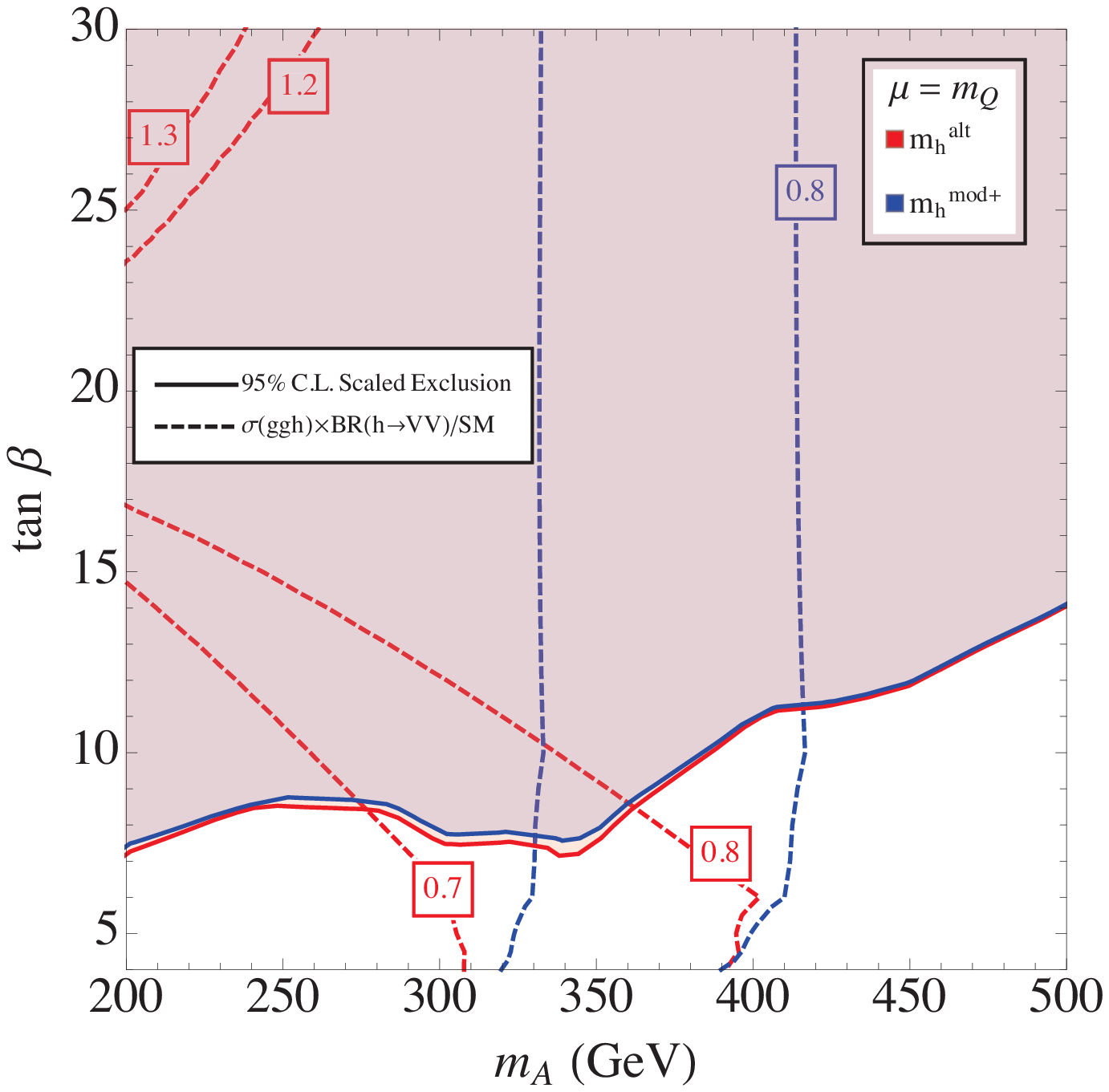}}~
\subfloat[]{\includegraphics[width=3.3in, angle=0]{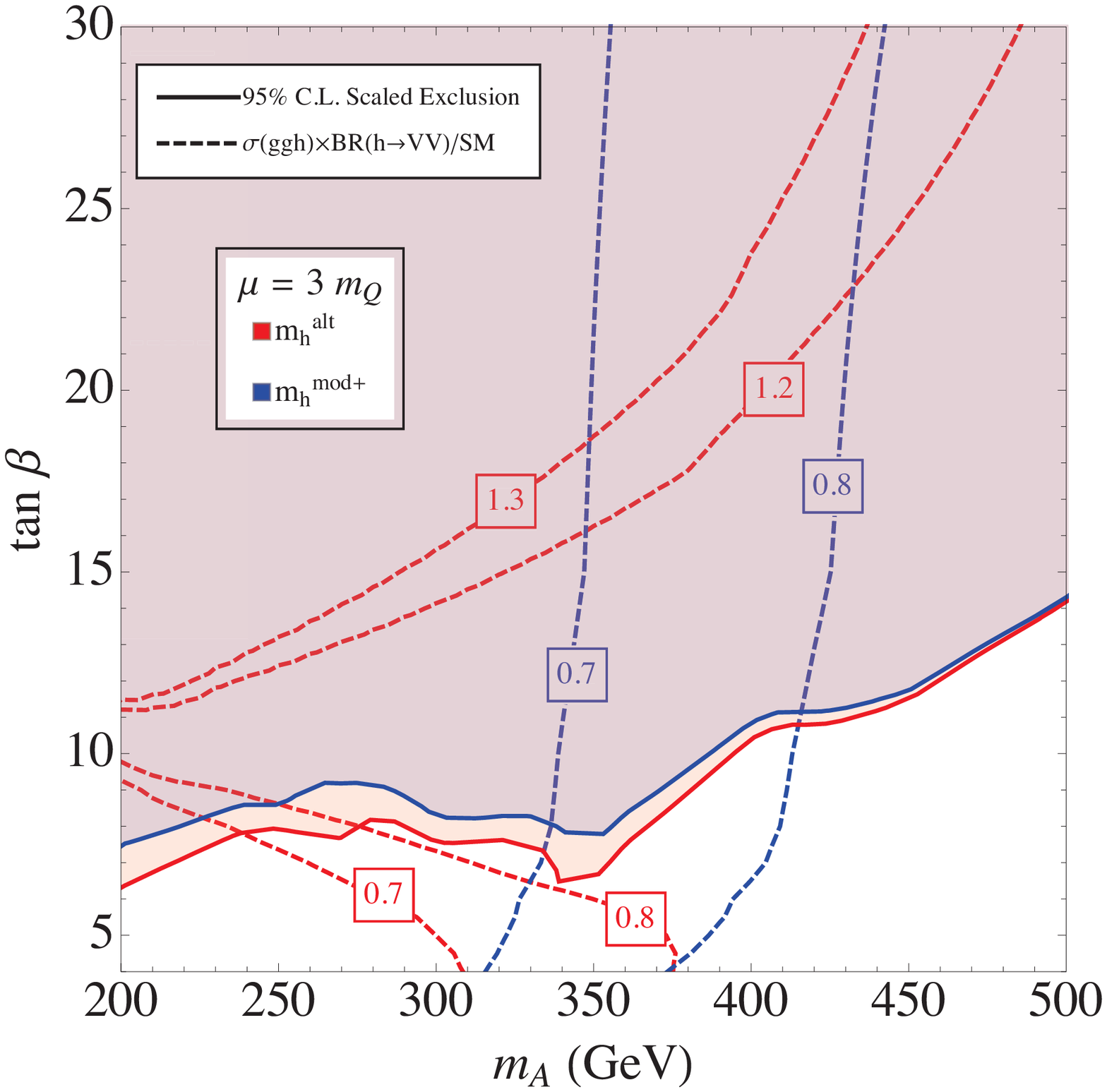}}
\caption{\label{fig:Overlaidbounds}{\em Dashed contours show deviations of the signal strength into massive gauge bosons for the lightest CP-even Higgs boson with respect to the SM values in the $m_h^{\rm mod +}$~(blue) and $m_h^{\rm alt}$~(red) scenarios in the $m_A$--$t_\beta$ plane for different values of $\mu$. Shaded regions denote parameters excluded by direct searches for heavy CP-even and CP-odd Higgs bosons decaying into pairs of $\tau$ leptons. }}
\end{figure}

In Fig.~\ref{fig:Overlaidbounds}, we summarize our results on the comparison of direct searches for non-standard Higgs bosons and the precision studies of the lightest CP-even Higgs boson at the 8~TeV LHC.  The dashed contours correspond to various assumptions on the precision of the signal strength $\sigma(gg\to h)\times {\rm BR}(h\to VV)$. For example, the $0.8$ contour corresponds to a signal strength that is $80\%$ of the predicted SM value, etc. The four panels represent four different values for the $\mu$ parameter, and in each panel we depict both the $m_{h}^{\rm mod+}$ and the $m_{h}^{\rm alt}$ scenarios. At low values of $\mu$, in light of the weak dependence of the light CP-even Higgs decay branching ratios  on $t_\beta$, precision studies of the decay branching ratios of the lightest CP-even Higgs lead to a lower bound on the value of $m_A$, which is roughly independent of $t_\beta$. Indeed, the dashed contours in Fig.~\ref{fig:Overlaidbounds}~(a) are nearly vertical, ruling out the parameter space to the left of the corresponding contours. The ATLAS experiment has performed such an analysis and found a bound on $m_A$ of order 400~GeV. Let us remark in passing that the signal strength of the $h \to VV$ modes observed at ATLAS is $1.3 \pm 0.2$ and hence according to the results of
Fig.~\ref{fig:Overlaidbounds}~(a) the bound on $m_A$ would be larger than the corresponding one using CMS data, for  which the signal strength is  $1.0 \pm 0.2$.

As the value of $\mu$ is increased we see two effects. On one hand, the contours of constant $h$-induced $VV$ production cross section are drastically modified in the $m_{h}^{\rm alt}$ scenario, due to a relevant dependence on $t_\beta$ of the bottom quark and tau lepton Yukawa couplings [cf.~\eqs{tanbcbma}{ghdd}]. These contours bend to the left in relation with the ones in the $m_{h}^{{\rm mod}+}$ scenario, becoming almost independent of $m_A$ at values of $t_\beta$ close to the alignment limit.  Therefore, for $t_\beta$ close to the value where the alignment condition is satisfied, precision measurements alone are not able to place any bound on $m_A$. The smallest value of $t_\beta$ where the alignment condition is satisfied takes place for the largest value of $\mu=3m_Q$ considered, shown in Fig.~\ref{fig:Overlaidbounds}~(d).\footnote{In Fig.~\ref{fig:Overlaidbounds}~(d), we have suppressed additional dashed red contour lines that reappear in the $m_h^{\rm alt}$ scenario with $\mu=3m_Q$ in the large $t_\beta$, low $m_A$ parameter regime.  In this regime, the magnitude of the $hb\bar{b}$ coupling is once again SM-like, but its sign is flipped relative to that of the $hVV$ coupling.  This wrong-sign $hb\bar{b}$ coupling regime, discussed in detail in Ref.~\cite{Ferreira:2014naa}, cannot be ruled out by the present $h(125)$ data alone, but is completely incompatible with the limits on the $H$ and $A$ direct searches via the $\tau^+\tau^-$ channel.} Indeed it is difficult to obtain smaller values of $t_\beta$ at alignment in the MSSM without taking extreme values of the MSSM parameters. Large values of $A_t/m_Q$ and $\mu/m_Q$ can lead to charge and color breaking vacua which would bring the stability of the electroweak vacuum into question~\cite{VacStab}.

The complementarity of the precision $h(125)$ data with direct searches for non-standard Higgs bosons is now clear. At the large values of $t_\beta$ where the alignment condition is satisfied, searches for non-standard Higgs bosons become effective and, as discussed in the previous section, they become more effective for larger values of $\mu$. This is shown by the shaded regions of Fig.~\ref{fig:Overlaidbounds}, which denote the CMS limits in the $m_{h}^{\rm mod+}$ and $m_{h}^{\rm alt}$ scenarios. The combination of direct and indirect searches allow us to constrain values of $m_A$ lower than 250~GeV in the $m_h^{\rm alt}$ scenario with $\mu \simlt 3 m_Q$, independently of $t_\beta$. Moreover, due to the increase in sensitivity of the search for non-standard Higgs bosons at large values of $\mu$, the whole region of parameters for $m_A < 350$~GeV is expected to be probed by the LHC in the near  future, showing again the strong complementarity between precision studies of the lightest CP-even Higgs boson, which become a weaker probe in this scenario, and  direct searches for non-standard Higgs bosons.

In summary, at low values of $\mu$, precision measurements of the lightest CP-even Higgs bosons are able to probe low values of $m_A$, independently of $t_\beta$. In contrast, in the presence of alignment which occurs for large values of $\mu$, precision measurement studies alone will not be able to put a model independent bound on $m_A$. However, in this case direct searches for non-standard Higgs bosons will be able to probe all values of $t_\beta$ for values of $m_A$ below the top-quark decay threshold in the near future.

\section{Conclusions}
\label{sect:6}

In this article, we have analyzed the complementarity between precision measurements of the lightest CP-even Higgs boson and direct searches for non-standard Higgs bosons in the MSSM.  We have stressed that in the alignment limit, one can significantly  relax the bounds on the heavy Higgs bosons that arise from the measurements of the $VV$ decays of the lightest CP-even Higgs boson.  Such alignment conditions, however, are associated with large values of the $\mu$ parameter and the stop mixing parameter, $A_t$, and tend to be restricted to values of $t_\beta$ of order 10 or larger within the MSSM.

Direct searches for non-standard neutral Higgs bosons provide strong constraints on the Higgs spectrum. Currently, the most sensitive search channel is associated with the $\tau^+\tau^-$ final state, with the main production mode being
either through the gluon fusion process or in association with bottom quarks.  The ATLAS and CMS experiments have placed lower bounds on $m_A$ that range from values of order 200~GeV for $t_\beta \simeq 10$ up to values of order of a TeV for $t_\beta \simeq 50$.  The lower values of $m_A$ and $t_\beta$ may be consistent with the observed lightest CP-even Higgs properties, provided one is not far from the alignment condition. The large values of $\mu$ associated with the alignment limit reduce the decay rate into charginos and neutralinos and therefore increase the BR($H,A \to \tau^+\tau^-)$, making direct searches more efficient. This property provides an interesting complementarity between direct searches and precision measurements which will allow one to probe the region of $m_A < 350$~GeV for all values of $t_\beta$ in future running of the LHC.

\begin{acknowledgements}
We gratefully acknowledge informative discussions with Greg Landsberg.
Fermilab is operated by Fermi Research Alliance, LLC under Contract No. DE-AC02-07CH11359 with the U.S. Department of Energy. University of Chicago is supported in part by U.S. Department of Energy grant number DE-FG02-13ER41958. H.E.H. is supported in part by U.S. Department of Energy grant number DE-FG02-04ER41286. Work at Northwestern is supported in part by the U.S. Department of Energy under Contract No. DE-SC0010143. Work at ANL is supported in part by the U.S. Department of Energy under Contract No. DE-AC02-06CH11357. N.R.S is supported by the U.S. Department of Energy grant No. DE-SC0007859 and by the Michigan Center for Theoretical Physics. M.C, H.E.H., N.R.S. and C.W  thank the hospitality of the Aspen Center for Physics, which is supported by the National Science Foundation under Grant No. PHYS-1066293. M.C., N.R.S and C.W also thank the hospitality of KITP, which is supported by the National Science Foundation under Grant No. NSF PHY11-25915.
 \end{acknowledgements}

\bigskip\bigskip\bigskip

\appendix
 \section{Interpreting Current Bounds from LHC8}
\label{sect:bounds}

In Ref.~\cite{Khachatryan:2014wca} where CMS presented bounds on the heavy Higgs bosons in the MSSM, the limits were derived in particular benchmarks that differ from the two classes of scenarios we are considering in this work. As such, these limits cannot be applied in a straightforward manner. However, Ref.~\cite{Khachatryan:2014wca} also provided {\em model-independent} limits that could be translated into limits in  benchmarks considered in this study. The model-independent limits are provided as two-dimensional contours in the plane of the production cross-sections via gluon fusion and associated production with bottom quarks. These limits are derived from searching for a heavy scalar resonance in the $\tau^+\tau^-$ final state, independently of any specific model, and show  very little contamination from a 125 GeV Higgs boson once the postulated heavy resonance is heavier than 200 GeV.

Unlike the model-independent bounds, the limits in the MSSM benchmarks in Ref.~\cite{Khachatryan:2014wca} are given in terms of $m_A$ and $t_\beta$, instead of direct upper bounds on the $\tau^+\tau^-$ production rates. We will specifically use the exclusions presented for the $m_h^{\rm mod+}$ scenario, compare them to the limits presented in the model independent analysis and formulate an algorithm to apply these to any other MSSM model. To that end, we first derive the upper limit on the production rates in the $m_h^{\rm mod+}$ scenario, with $\mu=200$ GeV, by computing the corresponding branching ratios and relevant cross-sections along the exclusion curve in the $m_A$--$t_\beta$ plane using the package {\tt FeynHiggs}~\cite{FeynHiggs}. For each value of $m_A$ there exists an upper limit on the allowed inclusive production rate into $\tau^+\tau^-$.  We will refer to this upper limit as the \textit{inclusive interpretation} of the heavy Higgs boson search bounds.

\begin{figure}[ht]
\subfloat[]{\includegraphics[width=3.3in, angle=0]{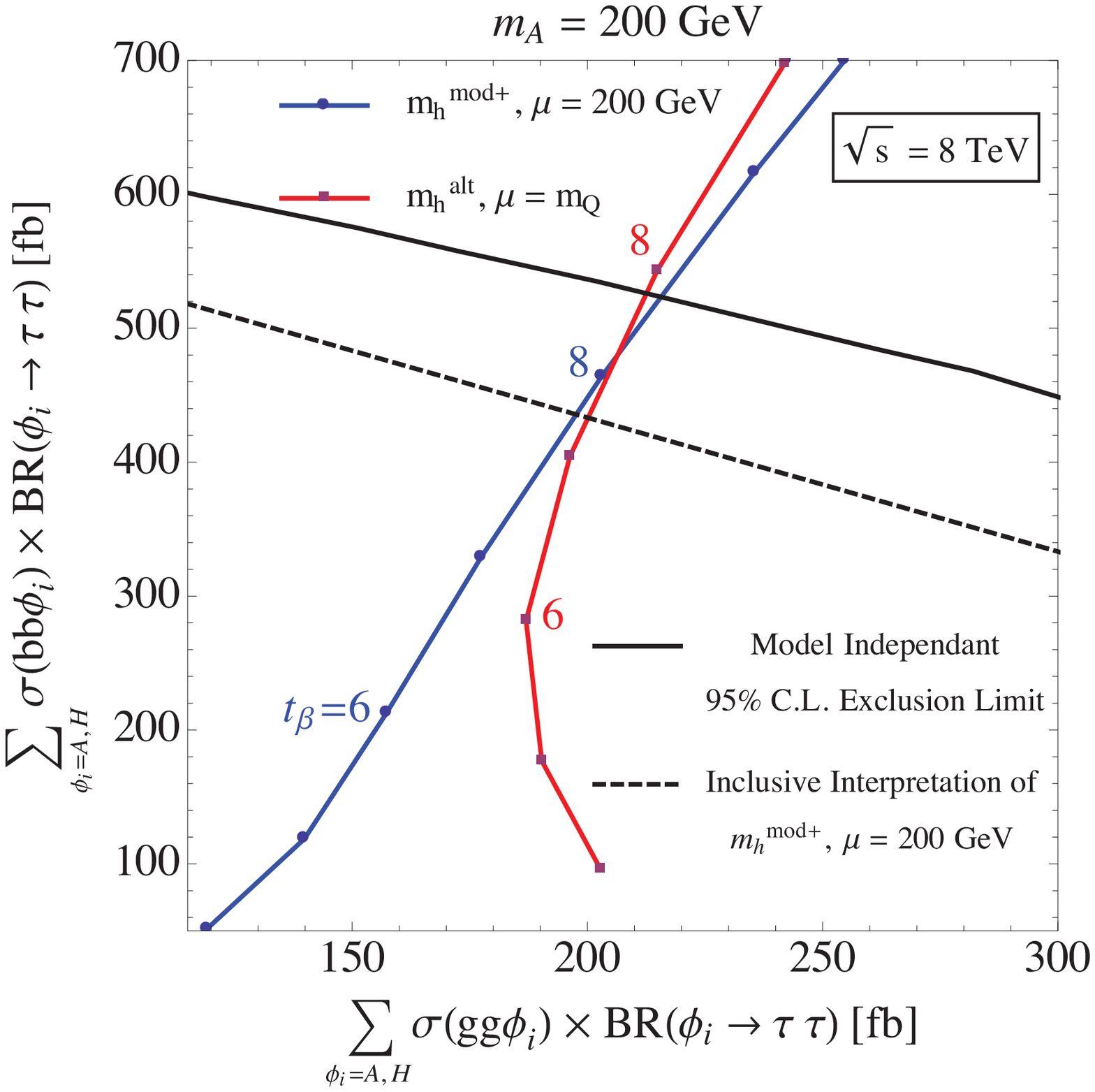}}~~
\subfloat[]{\includegraphics[width=3.3in, angle=0]{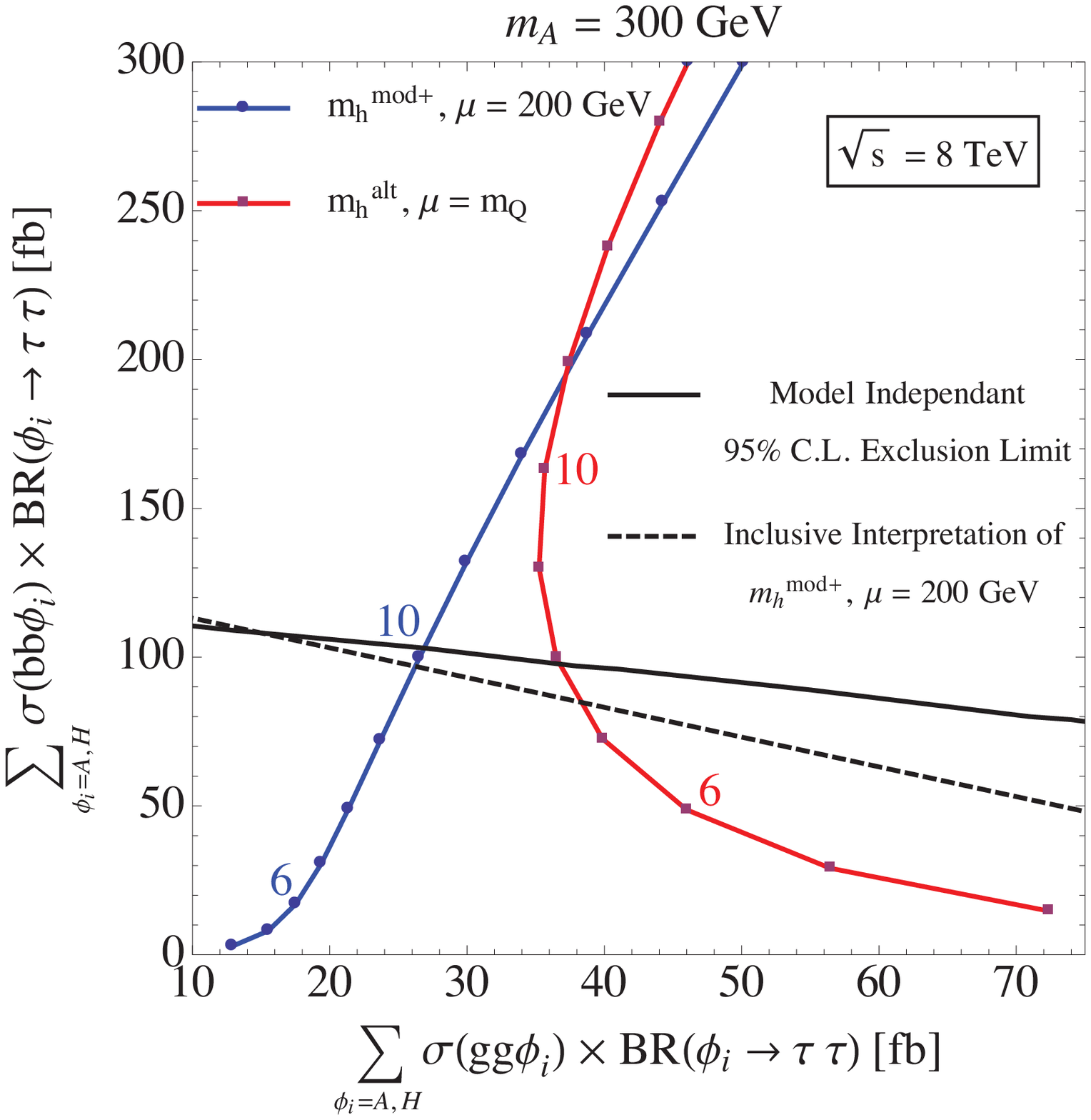}}
\caption{\label{fig:bbvsgg}{\em Comparison of exclusion limits obtained via the model independent  analysis and our inclusive interpretation of the limits for the $m_h^{\rm mod+}$ scenario, with $\mu=200 \;\mathrm{GeV}$. The dots represent values of $t_\beta$ in units of 1, where values are labeled in blue or red corresponding to the $m_h^{\rm mod+}$ scenario, with $\mu=200\;\mathrm{GeV}$, and $m_h^{\rm alt}$scenario, with $\mu= m_Q$,  respectively. }}
\end{figure}

In Fig.~\ref{fig:bbvsgg} we show the production rates into $\tau^+\tau^-$ resulting from the production of heavy Higgs bosons in the two relevant production channels, $gg\phi$ and $bb\phi$, as a function of $t_\beta$. The production
 rates in the $m_{h}^{\rm alt}$ scenario, for $\mu = m_Q$,  are displayed as a solid red curve, while the corresponding values in the $m_{h}^{\rm mod+}$ scenario, for $\mu = 200$~GeV, are displayed as a solid blue curve. We show results for $m_A = 200$~GeV and $m_A=300$~GeV in the left and right panels of Fig.~\ref{fig:bbvsgg}, respectively. The corresponding values of $t_\beta$ are displayed as solid dots on these curves.  These show that, while in the $m_{h}^{\rm mod+}$ scenario, with $\mu=200$ GeV,  the rates due to both production cross sections decrease with $t_\beta$, the rate originating from the production via gluon fusion reaches a minimum in the $m_{h}^{\rm alt}$ scenario, increasing at low values of $t_\beta$ in agreement with our discussion in Sec. III C.

Our inclusive interpretation of the heavy Higgs boson search is denoted by dashed black lines in  Fig.~\ref{fig:bbvsgg}. We also show the model-independent 95\% C.L. upper bounds, provided explicitly in Ref.~\cite{Khachatryan:2014wca}, as black solid lines.  Observe that the slopes of the solid and dashed lines are very similar, implying that the model independent bounds correspond approximately to the same inclusive production rate in both scenarios. Note the bound on $t_\beta$ we obtain in the m$_h^{\rm mod+}$ scenario, with $\mu=200$ GeV,  from the model-independent bounds is within one unit of the bound presented by CMS by a more sophisticated likelihood method.

The $t_\beta$ limit for a given $m_A$ in a different MSSM model corresponds roughly to the value where the inclusive production rate exceeds the upper limit in the $m_h^{mod+}$ scenario, with $\mu=200$ GeV.  Since the sensitivity of the LHC in the gluon fusion and $bb\phi$ channels is similar, we expect this to be a good approximation. Explicitly, in Fig.~\ref{fig:bbvsgg} we show the comparison of the bound in the m$_h^{\rm alt}$ scenario, with $\mu=m_Q$, using the inclusive production rate at the limiting value of $t_\beta$ presented by CMS in the m$_h^{\rm mod}$ scenario, with $\mu = 200$~GeV, compared to the limit on the value of $t_\beta$ that could be interpreted from the model independent bound. Again, the difference using the two methods results in a difference for the $t_\beta$ limit of approximately one unit.

Using our inclusive interpretation, we can scale the limits from the $m_h^{\rm mod+}$ scenario, with $\mu = 200$~GeV, to any other scenario in a simple way in the region where $m_A$=200--350 GeV. We then use the bounds from our inclusive interpretation to map out the direct search constraints on the $m_A$--$t_\beta$ plane in each of our benchmarks, which in turn are compared against the constraints from precision measurements of the properties of the 125 GeV Higgs boson.  We also use the inclusive production rate to analyze the future searches at the 14~TeV run of the LHC.

\section{Comparison of $\boldsymbol{hVV}$ and $\boldsymbol{h\gamma\gamma}$ Couplings}
\label{appen}

At low values of $\mu$ the charginos become light and therefore can lead to a modified diphoton coupling of the lightest CP-even Higgs boson. The contribution of stops and charginos to the amplitude in the diphoton channel is proportional to~\cite{Ellis:1975ap,Shifman:1979eb,Gunion:1989we,Kniehl:1995tn,CLW}
\begin{equation}
\mathcal{A}_{h\gamma\gamma}  \simeq {\cal A}_{h\gamma\gamma}^{\rm SM} +
b_{\tilde{\chi}^+} \frac{ \half g^2 v^2 \sin 2\beta}{ M_2  \mu - \frac{1}{4} g^2 v^2 \sin 2 \beta} - b_{\tilde{t}} \ m_t^2 \frac{ m_{\tilde{t}_1}^2 + m_{\tilde{t}_2}^2 - X_t^2}{m_{\tilde{t}_1}^2 m_{\tilde{t}_2}^2 }\;,
\end{equation}
where in this normalization $\mathcal{A}_{ h \gamma \gamma}^{\rm SM} = 6.5$ represents the SM contribution, $b_{\tilde{\chi}^+} = 4/3$,  $b_{\tilde{t}} = 4/9$, and $m_{\tilde{t}_{1,2}}$ are the stop mass eigenvalues. The parameters $m_t$ and $X_t$ are running mass parameters at the scale of the stop masses in the $\overline{\rm MS}$ scheme.  For the large values of $X_t$ present in the $m_{h}^{\rm alt}$ scenario, the stop contribution is small and positive. The chargino contribution is also small, and becomes only relevant for small values of $\mu$ and of $t_\beta$.   In the $m_{h}^{\rm mod+}$ scenario, for $\mu = 200$~GeV, the stop contribution is even smaller, since $X_t^2$ is close to the sum of the squares of the stop masses.  In general, the supersymmetric loop corrections lead to a contribution of the order of a few percent of the SM one.  Hence, the main deviation of the BR$(h \to \gamma \gamma)$ and BR$(h \to VV)$ in this region of parameters is mostly governed by the increase of the width of the lightest CP-even Higgs decay into bottom quarks and tau leptons at low values of $m_A$. 

Note that the contribution from stops to gluon fusion is approximately a factor of 3 larger than their contribution to the diphoton coupling~\cite{Kniehl:1995tn,CLW,Carena:2013iba}. However, the leading SM contribution has the opposite sign in this case, and hence, the gluon fusion rate is reduced from the SM expectation in the scenarios we consider, again at the few percent level.

\begin{figure}[t]
\subfloat[]{\includegraphics[width=3.3in, angle=0]{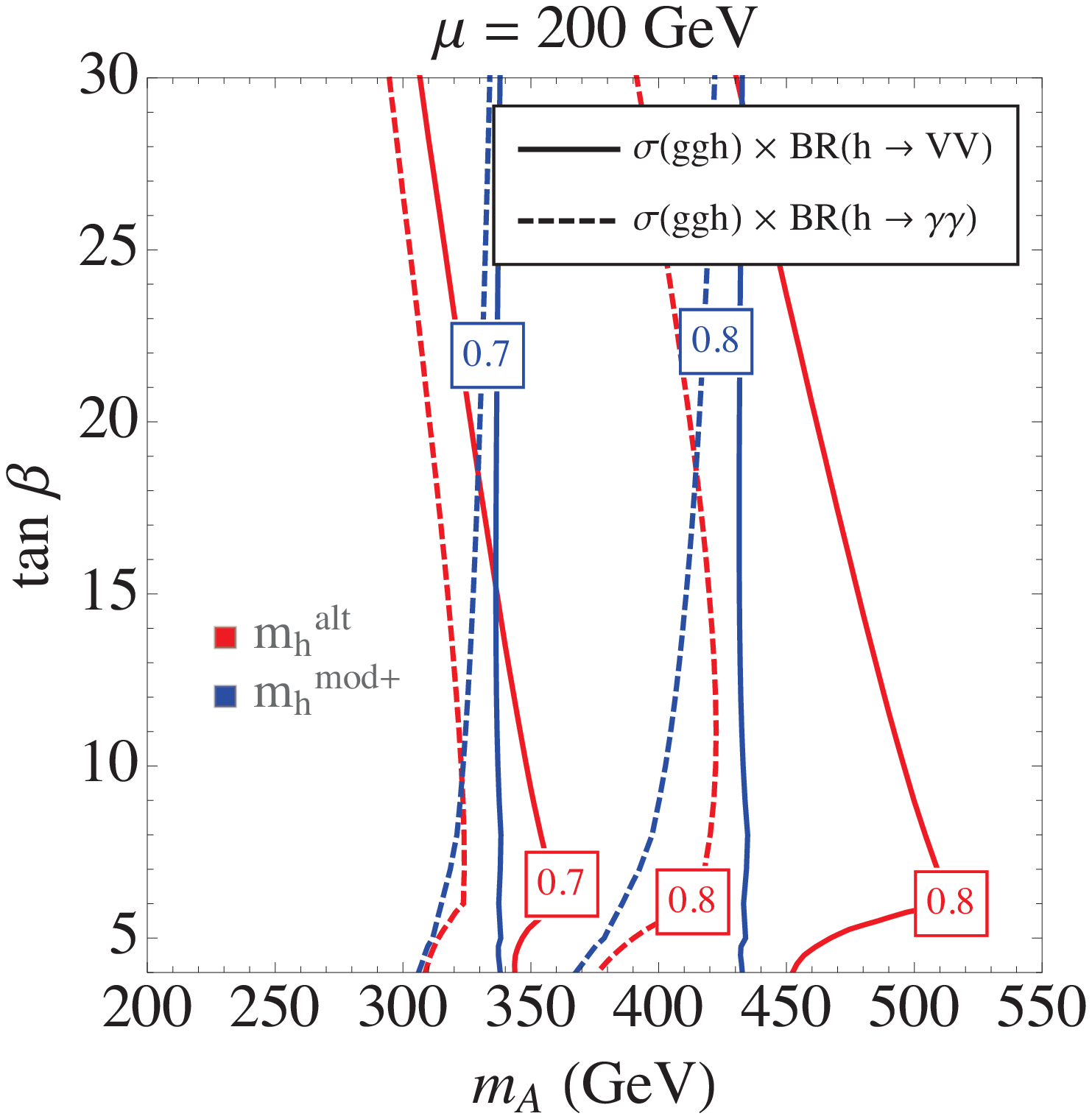}}~
\subfloat[]{\includegraphics[width=3.3in, angle=0]{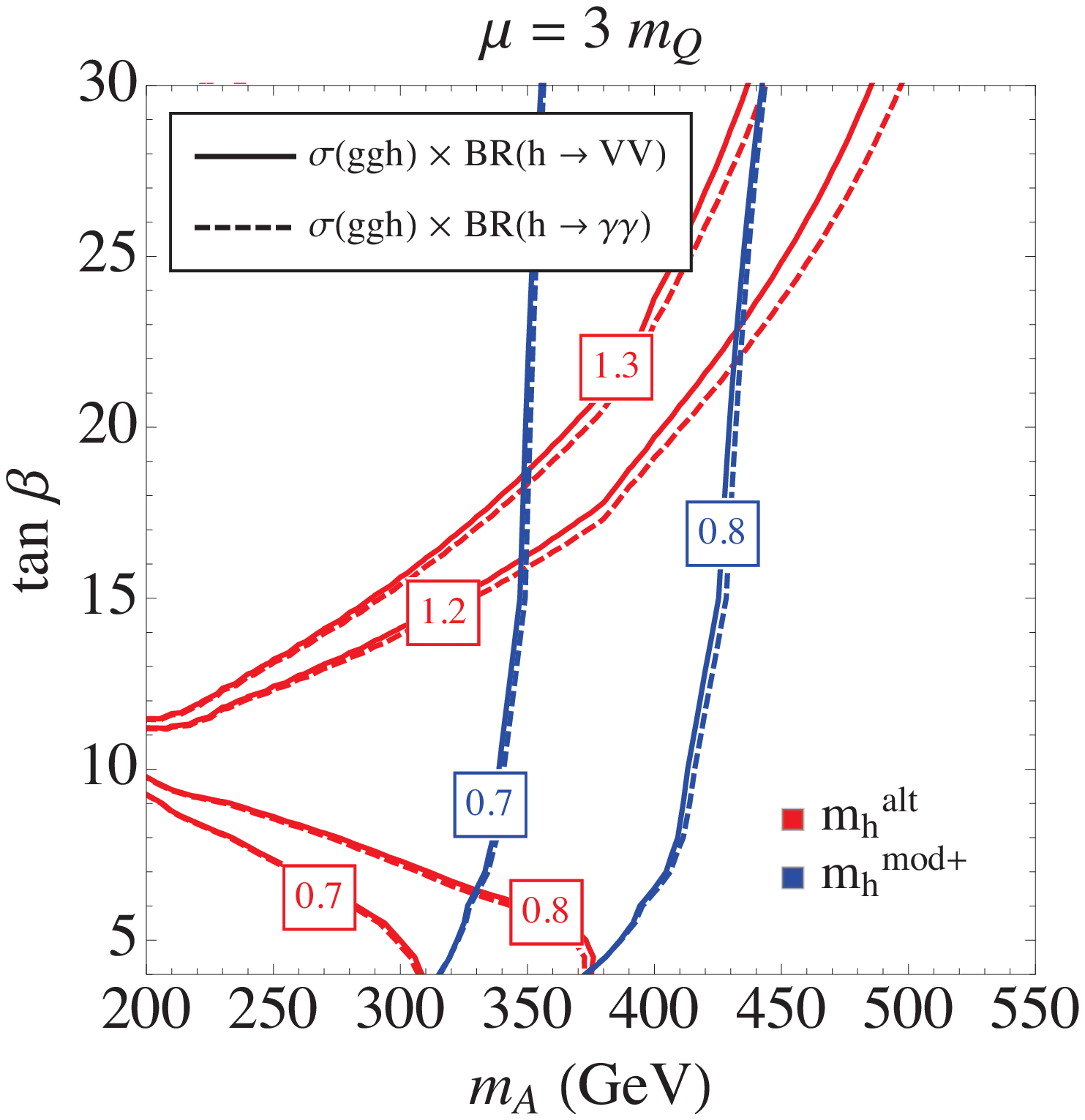}}
\caption{\label{fig:WWvsgaga}{\em Deviation of the signal strengths with respect to the SM values for the lightest Higgs boson decaying into two photons and two massive gauge bosons.}}
\end{figure}

In order to quantify these effects, in Fig.~\ref{fig:WWvsgaga}~(a) we show contour plots of $\sigma \times$~BR($h\to \gamma\gamma$) and $\sigma \times$~BR($h\to VV$)  normalized the the SM values in the $m_{h}^{\rm mod+}$ and $m_{h}^{\rm alt}$ scenarios for low values of $\mu$, for which no alignment condition is present.  This choice of $\mu$ maximizes the differences between these channels. As can be seen, the overall behavior of these channels is the same, although the precise value of $t_\beta$  for which a particular deviation with respect to the SM value takes place is shifted by a few tens of GeV for $m_A < 350$~GeV for low values of $\mu$. No significant difference is present for larger values of $\mu$, as can be seen from Fig.~\ref{fig:WWvsgaga}~(b). The peculiar behavior of the contour lines at low values of $t_\beta$ in the $m_{\rm h}^{\rm alt}$ scenario is induced by the variation of the gluon fusion cross section, which becomes more suppressed as the stops become heavier.

In this article, in order to study the properties of the lightest CP-even Higgs bosons we shall concentrate on the BR$(h \to VV)$, but as shown in Fig.~\ref{fig:WWvsgaga}, similar conclusions would be obtained by the study of BR$(h \to \gamma\gamma$) in this region of parameters.

\end{document}